\documentclass[11pt,a4paper]{article}
\usepackage[utf8]{inputenc}
\usepackage{a4}
\usepackage{amsmath,amssymb}
\usepackage[labelfont=bf]{caption}
\usepackage{cite}
\usepackage{color}
\usepackage{mathtools}
\usepackage{graphicx}
\usepackage{silence}
\usepackage{siunitx}
\usepackage{subcaption}
\allowdisplaybreaks
\usepackage{hyperref}
\hypersetup{
    colorlinks=true,
    linkcolor=black,
    filecolor=black,
    urlcolor=black,
    citecolor=black
    }
    
\def\lin{\lambda_{\rm in}}
\def\dgeo{d_{\rm geom}}
\def\daero{d_{\rm aero}}
\def\dstokes{d_{\rm S}}
\def\dstokesi{d_{\text{S},i}}
\def\elpi{ELPI+}
\def\rhop{\rho_{\rm part}}
\def\rhopi{\rho_{\text{part},i}}
\def\rhoas{\rho_{\rm aerosol}}
\def\jcal{J_{\rm cal}}
\def\jjet{J_{\rm imp}}
\def\xdil{X_{\rm dil}}
\def\K{K_{\rm e}}

\newcommand{\defeq}{\vcentcolon=}
\newcommand{\eqdef}{=\vcentcolon}

\begin{document}
\thispagestyle{empty}

\begin{center}
{\huge\bf
Extinction coefficients\\[.2cm]
from aerosol measurements}

\vspace{4em}
{\sc\large
Christoph Gnendiger$^{a}$,
Thorsten Schultze$^{b}$,\\[.1cm]
Kristian Börger$^{a,c}$,
Alexander Belt$^{a}$,
Lukas Arnold$^{a,c,}\footnote{e-mail:
    \href{mailto:l.arnold@fz-juelich.de}{l.arnold@fz-juelich.de},\,%
    \href{mailto:arnold@uni-wuppertal.de}{arnold@uni-wuppertal.de}}$
}\\[2em]
{\sl
${}^a$Institute for Advanced Simulation (IAS),\\
Forschungszentrum Jülich,\\
D-52428 Jülich, Germany\\
\vspace{0.3cm}
${}^b$Chair of Communication Systems (NTS),\\
University of Duisburg‐Essen,\\
D-47057 Duisburg, Germany\\
\vspace{0.3cm}
${}^c$Chair of Computational Civil Engineering (CCE),\\
University of Wuppertal,\\
D-42119 Wuppertal, Germany\\
}
\end{center}
\vspace{4ex}

\begin{abstract}
In this contribution, we develop a model based on
classical electrodynamics that describes
light extinction in the presence of arbitrary aerosols.
We do this by combining aerosol and light-intensity
measurements performed with the well-proven
measuring systems \elpi\ and MIREX, respectively.
The developed model is particularly simple and
depends on only a few input parameters, namely
on densities and refractive indices of the
constituting aerosol particles.
As proof of principle, the model is in first
applications used to determine extinction coefficients
as well as mass-specific extinction for an infrared light source
with a peak wave length of $\lambda\!=\!\qty{0.88}{\text{\textmu}\m}$.
In doing so, detailed studies concentrate on two
aerosols exemplary for characteristic values of the
input parameters:
a non-absorbing paraffin aerosol in a bench-scale
setup and soot from a flaming $n$-heptane fire
in a room-scale setup
(test fire TF5 according to standard EN54).
As main results, we find numerical values for mass-specific
extinction that are first of all different in the
two considered cases.
Moreover, obtained results differ in part more than a factor
of three from literature values typically used in practical
applications.
Based on the developed model, we explicitly address and
assess underlying reasons for the deviations found.
Finally, we propose a simple way how future light-extinction
studies can be performed comparatively easily by means of the
\elpi-system or measuring devices that work in a similar way.
\end{abstract}


\newpage
\noindent\hrulefill
\vspace*{-10pt}
\tableofcontents
\noindent\hrulefill

\section{Introduction}

Visibility in case of fire is an essential
component in the performance-based safety design of
buildings and therefore an indispensable aspect in the
field of civil safety applications and research.
Accordingly, a multitude of experiments investigating
the human capacity to identify safety-relevant objects
in the presence of soot has been conducted in the past,
the first experiments dating back to the early 70s
of the last century
\cite{Jin_1970,Jin_1971,Jin_1972,Jin_1973,Jin_1978}.
While the application field of~the gained knowledge was
initially limited to manual engineering procedures,
it is nowadays also frequently applied
in the field of computational fluid dynamics (CFD),
for instance in the widely used
\textit{Fire Dynamics Simulator} FDS \cite{fds}.
A reliable and comprehensive description of visibility 
in the presence of an aerosol has therefore
become of central importance for the safety
assessment of buildings and other places
where people are present.

The key quantity to describe visibility in the
context of environmental conditions is the
\textit{extinction coefficient} which quantifies
properties of an aerosol to attenuate parts of a
light beam from its original path.
Conceptually, light extinction can be measured in two
different ways:
In a first approach, the focus is put on the light beam
and the total amount of light-intensity reduction
compared to the case without aerosol is measured.
Corresponding experiments are often called `direct'
visibility measurements and are usually evaluated
using the Beer–Lambert–Bouguer law~\cite{bohren2009}.
In contrast, in a second category of experiments,
the focus is put on the aerosol
instead and, for instance, quantities like
number and mass densities, sizes, or even fractal properties
of the constituting aerosol particles are measured.
Using models that typically rely on Rayleigh or Mie
computations and extensions thereof,
visibility reduction is then deduced from
the measured particle properties.
Since in the latter approach visibility conditions are
derived only in a downstream step and in the framework
of the applied model, these measurements are sometimes
denoted as `indirect' ones.

Historically, there has been considerable restraint
and criticism regarding the 'indirect' approach.
Objects of concern typically comprise
the alleged presence of large experimental
uncertainties~\cite{
Dobbins1994ComparisonOA,
Widmann2003MeasurementOT}
or the used models for light interaction of the
aerosol particles~\cite{
Dobbins1994ComparisonOA,
Wu1997,
Widmann2003MeasurementOT}.
Accordingly, many statements can be found in the literature
claiming that `direct' visibility measurements are
the only reliable, or at least the preferable way to
determine optical properties of aerosols.
An interesting quantity in this respect is the
\textit{mass-specific} extinction coefficient.
Since it combines, by definition,
light extinction with the overall mass density
of the present aerosol, mass-specific extinction can not be
obtained through `direct' visibility measurements alone
but has always to be combined with other experimental
results like gravimetric ones.
In the literature, the existence of a (universal)
mass-specific extinction coefficient
has so far been investigated empirically,
based on the hypothesis that post-flame smoke
produced by overventilated fires behaves almost
identically and is more or less independent of
the burned fuel
(see for instance the review in \cite{Widmann_2003}
and references therein).
Although functional dependences on different
influencing factors like the wave length of
the incident light \cite{Widmann_2003} or the
ventilation state of the fire
\cite{Widmann2003MeasurementOT} are qualitatively known,
these aspects are often neglected for the sake of simplicity
and, accordingly, the same constant value for mass-specific
extinction is used in practical applications.
FDS, for instance, uses by default the constant
value proposed in \cite{Mulholland_Croarkin_2000}
which itself is the review result of various
small-scale experiments that have been conducted
under well-controlled conditions.
Given this fact, it is of major importance to know
if universal mass-specific extinction does exist
at all and, if so, to know its actual value in
real-scale applications.
Only then, it is possible to draw reliable
conclusions that neither overestimate visibility
conditions (which would be detrimental for safety
aspects) nor underestimate them (which could result
in unnecessary high expenses for safety measures).

In the present contribution, we therefore put
the determination of (mass-specific) extinction
on a firm basis by combining
`direct'  and `indirect' visibility measurements.
More precisely, we combine aerosol measurements
performed with the cascade impactor \elpi~%
\cite{elpi} and light-intensity reduction measurements
performed with the extinction measuring equipment
MIREX~\cite{mirex}.
The goal of the comparison is to develop
a practical model for the interaction of
a light beam with aerosol particles that
\begin{itemize}
    \item allows to reliably predict the
        total amount of light-intensity reduction\\
        in the presence of an arbitrary aerosol,
    \item is at the same time compatible with
        known value ranges\\
        of the entering model parameters,
    \item has expected predictive uncertainties\\
        comparable with `direct' visibility measurements, and
    \item is preferably as simple as possible.
\end{itemize}
In order to achieve this, the structure of the
paper is as follows:
In Secs.\,\ref{sec:soot_measurements} and
\ref{sec:visibility_measurements}, we summarize
key aspects of \elpi- and MIREX-measurements
in general and identify possible
output of the two systems allowing
to describe light extinction in a
combined framework.
Building on these findings, in
Sec.\,\ref{sec:theoretical_framework}
we develop a first-principle model
that is based on classical electrodynamics and that
fulfils the above requirements.
In doing so, we explicitly assess all relevant
assumptions made in the formulation of the model and
discuss possible modifications that may
(or may not) improve the predictive power in
practical applications.
In Sec.\,\ref{sec:results}, we then apply the model
and consider two different aerosols,
namely paraffin and soot from flaming
$n$-heptane fires which are exemplary for
characteristic differences of the input variables.
For the two cases, we determine value
combinations of the model parameters which allow to
bring \elpi- and MIREX-measurements into the best
possible agreement.
This is done by explicitly computing 
extinction coefficients in Sec.\,\ref{sec:ext_coeff}
and mass-specific extinction in
Sec.\,\ref{sec:specific_extinction}.
Before concluding, in Sec.\,\ref{sec:application}
we finally propose a simple methodology based
on the developed model that allows to comparatively
easily determine light extinction in future
applications of the \elpi-device.\\


\section{Measuring and modelling light interaction with an aerosol}

\subsection{Aerosol measurements with \elpi}
\label{sec:soot_measurements}

The \elpi-device is a cascade impactor which counts
and classifies aerosol particles according to their
aerodynamic and gravimetric properties \cite{elpi}.
Its working principle is as follows:
In a first step, a sample of the aerosol under investigation
is aspirated through a nozzle with an opening size of
$\qty{10}{\text{\textmu}\m}$ and the contained aerosol particles
are charged in a corona charger.
For each electrically charged particle, the total amount of
collected charge depends on the \textit{surface} of the
particle and therefore indirectly on its size and shape.
After charging, the aerosol sample is formed to a jet that
successively passes $14$ different impactor plates.
Before passing each of the plates, the particles have
to make a sharp turn, which for larger particles
is not possible due to larger inertia.
Accordingly, these particles are collected in the
respective impactor plate and the arising
electrical current~$I(t)$ is measured.
At the same time, smaller particles remain in the 
jet flow and move to the next impactor stage.
Since the opening size of the device is decreasing from
stage to stage, the speed of the jet flow is successively
increasing, allowing also smaller particles to be collected.
In this way, particles are collected and
classified from larger to smaller size.
The opening size of the lowest impactor stage is
$\qty{0.006}{\text{\textmu}\m}$.

As an output of an individual \elpi-measurement one
obtains in total $14$ time-dependent electrical currents
$I_i(t)$ with a time resolution of one second.
Using conversion factors, measured currents
can be converted into physical quantities like number
densities $\rho_{\text{N},i}$ and mass densities
$\rho_{\text{M},i}$ of the collected aerosol part
\cite{elpi}:
\begin{subequations}
\label{eq:elpi_densities}
\begin{align}
    \rho_{\text{N},i}(t) & = \frac{N_i(t)}{V_{\rm u}}\,
        = X_{N_i}\cdot I_{i}(t)\,,
        \phantom{\rho_{\text{N},i}(t)\cdot
            \rhopi\cdot}
        \qquad i\in\{1,\dots,14\}\,,
    \label{eq:elpi_densities_a}\\
    \rho_{\text{M},i}(t) & = \frac{M_i(t)}{V_{\rm u}}
        = \rho_{\text{N},i}(t)\cdot
            \left(\pi\frac{\dstokesi^3}{6}\right)\cdot
            \rhopi\,,
        \qquad i\in\{1,\dots,14\}\,.
    \label{eq:elpi_densities_b}
\end{align}
\end{subequations}
By convention,
the ordering of the channel labels is such that $i\!=\!1$
denotes the \elpi-channel collecting the smallest particles,
whereas $i\!=\!14$ corresponds to the channel
collecting the largest particles.
In Eqs.\,\eqref{eq:elpi_densities}, $N_i$ and $M_i$
denote absolute particle numbers and particle masses
per unit volume $V_{\rm u}\!=\!\qty{1}{\cm\cubed}$
of the measured aerosol and the conversion factors
$X_{N_i}$ consider charging and collection
inefficiencies of the \elpi-device.
The latter are calibrated in~%
\cite{Jaervinen_Aitomaa_Rostedt_Keskinen_Yli-Ojanperae_2013}
and provided in Appendix~A1 for the
sake of completeness.
Explicit definitions of $\dstokesi$ and $\rhopi$
are given in the next paragraph.

A crucial point in the evaluation of \elpi-data
is that each of the $14$ measured currents~$I_i$
(and therefore each of the derived densities) can be assigned
a dedicated particle size which is directly related
to aerodynamic and gravimetric properties of the collected
particles.
For that, each (irregularly shaped) aerosol particle is
internally converted into a virtual \textit{equivalent particle}
which is spherical and which has the same settling velocity~$v_s$
as the actual aerosol particle.
While the settling velocity of the aerosol particle
is indirectly measured by the \elpi-device, it is for the
virtual equivalent particle given by Stoke's law,
i.\,e.\
\begin{align}
    v_s =
        \frac{g}{18\,\mu_{\rm air}}
        (\rho-\rho_{\rm air})\cdot d^2\,.
    \label{eq:settling_velocity}
\end{align}
Here, $g\!\approx\!\qty{9.81}{\m\per\s\squared}$
is the standard acceleration due to gravity,
$\mu_{\rm air}\!\approx\!1.8\!\cdot\!10^{-5}%
\,\unit{\kg\per\m\per\s}$ and 
$\rho_{\rm air}\!\approx\!1.2\,\unit{\kg\per\m\cubed}$
are the dynamic viscosity and mass density of air
at a given temperature, respectively.
The three quantities are related to
measuring conditions of the experiments and are
approximately the same for each individual
\elpi-measurement.%
\footnote{Of course, $g$, $\mu_{\rm air}$, and
    $\rho_{\rm air}$ are affected by various
    boundary conditions of the experiments like
    ambient temperature or height above
    sea level.
    Compared to other sources of uncertainty,
    however, these factors can be regarded negligible,
    as will be shown below.
}
In contrast, $\rho$ and $d$ denote the mass density
and the diameter of the equivalent particle,
respectively.
The two quantities therefore strongly depend
on the measured aerosol.
For a given settling velocity, there are two 
common ways to \textit{define} the diameter of
the equivalent sphere, see also
Fig.\,\ref {fig:elpi_diameters}:

\begin{figure}[t!]
    \centering
    \begin{subfigure}{.70\textwidth}
        \includegraphics[width=1.0\linewidth]{%
            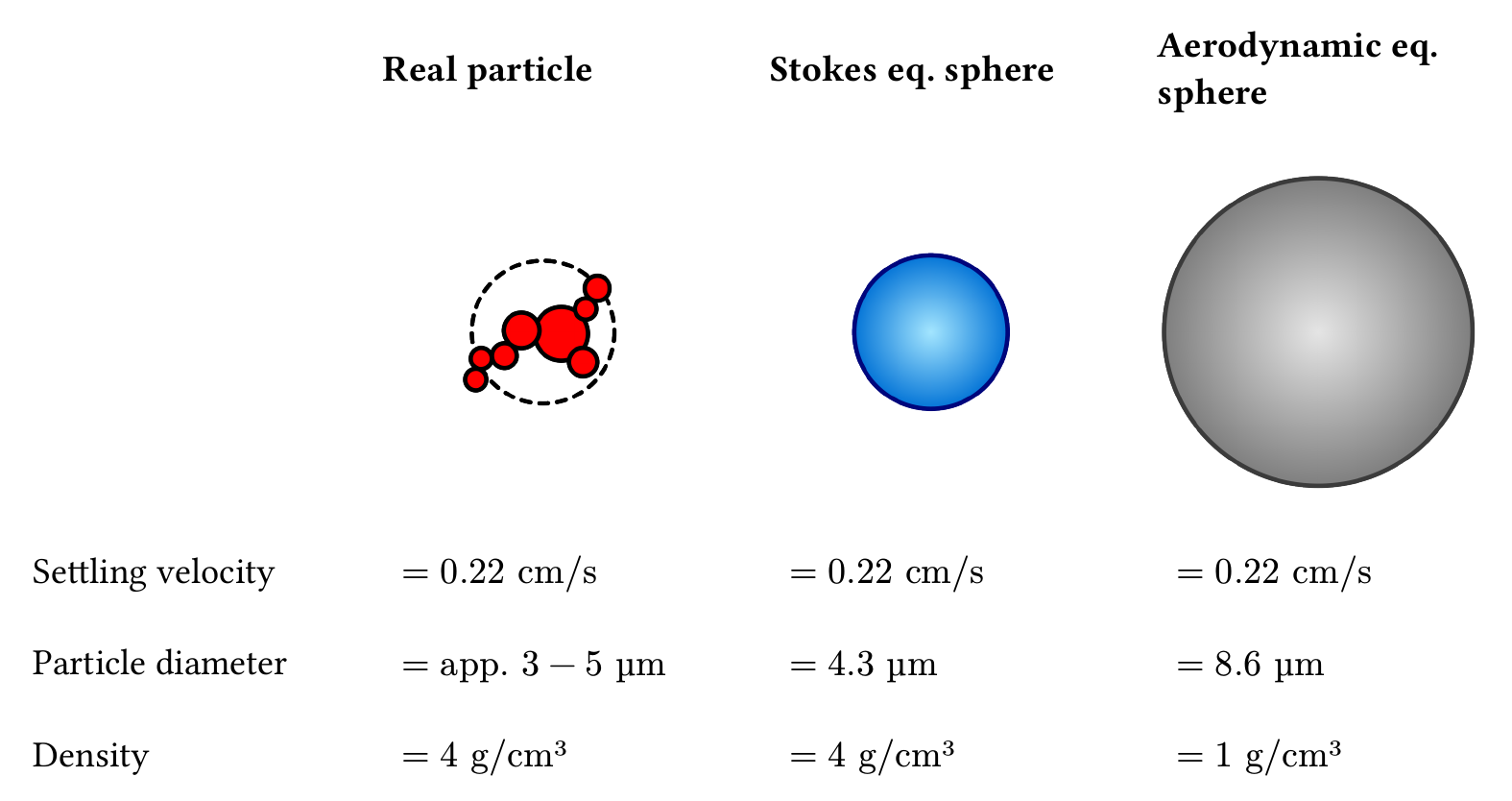}
    \end{subfigure}
    \caption{
        \textbf{Possible definitions of the equivalent diameter.}
        The figure is adopted from the \elpi-manual \cite{elpi}.
    \label{fig:elpi_diameters}}
\end{figure}

\begin{subequations}
\label{eq:diameters}
\begin{itemize}
    \item[(1)] \textbf{Stokes equivalent diameter $\dstokes$}\\
        In this case, the diameter of the equivalent particle
        is defined such that the latter has the same
        settling velocity and the same particle density
        $\rhop$ as the real aerosol particle:
        \begin{align}
             v_s \eqdef
            \frac{g}{18\,\mu_{\rm air}}
            (\rhop-\rho_{\rm air})\cdot\dstokes^2\,.
            \label{eq:settling_velocity1}
        \end{align}
    \item[(2)] \textbf{Aerodynamic equivalent diameter $\daero$}\\
        In this case, the diameter of the equivalent particle
        is defined such that the latter has the same
        settling velocity as the real aerosol particle.
        However, the equivalent particle is assumed to have
        unit density, i.\,e.\
        $\rho\!=\!\qty{1}{\g\per\cm\cubed}$:
        \begin{align}
             v_s \eqdef
            \frac{g}{18\,\mu_{\rm air}}
            (\qty{1}{\g\per\cm\cubed}-\rho_{\rm air})\cdot\daero^2\,.
            \label{eq:settling_velocity2}
        \end{align}
\end{itemize}
\end{subequations} 
It should be noted that both definitions introduce
auxiliary objects that may differ from the
actual size of the initial aerosol particle.
For instance, setting equal both definitions in
Eqs.\,\eqref{eq:diameters} and neglecting the
density of air, it is straightforward to derive
the following relation:%
\footnote{
    Actually, the functional dependence on the value
    of $\rhop$ is slightly more complicated and even
    depends on the respective \elpi-channel,
    see in particular the \elpi manual \cite{elpi}.
    However, Eq.\,\eqref{eq:ratio}
    serves as a useful approximation for the
    present considerations.
    }
\begin{align}
    \frac{\dstokes}{\daero}
    \approx\sqrt{\frac{\qty{1}{\g\per\cm\cubed}}{\rhop}}\,.
    \label{eq:ratio}
\end{align}
Accordingly, in case the actual particle density
$\rhop$ is larger (smaller) than
$\rho\!=\!\qty{1}{\g\per\cm\cubed}$,
aerodynamic diameters are larger (smaller) than
corresponding Stokes diameters.
Since an exemplary particle density of
$\rhop\!=\!\qty{4}{\g\per\cm\cubed}$ is assumed in
Fig.\,\ref {fig:elpi_diameters}, for instance,
the diameter of the aerodynamic equivalent sphere is
twice as large as the diameter of the corresponding
Stokes sphere and therefore much larger than the size
of the real aerosol particle.
In contrast, the given Stokes equivalent diameter
is compatible with the geometrical size of the
irregularly shaped aerosol particle
since it considers one additional 
physical property, namely the density $\rhop$ 
of the aerosol particle.
In order to get equivalent diameters
that are close to the geometries
of the aerosol particles, therefore
Definition~\eqref{eq:settling_velocity1} should be
used instead of \eqref{eq:settling_velocity2}.
In Appendix~A3 we further investigate
differences between both approaches and explicitly
quantify the impact of the choice on obtained
numerical results.
At this point it remains to be mentioned that,
since $\rhop$ is not measured by the \elpi-device,
its value has either to be measured separately or
to be taken from the literature.

Introducing Stokes equivalent spheres
via Eq.\,\eqref{eq:settling_velocity1}
allows to assign dedicated particle diameters
to each of the $14$ measured currents $I_i(t)$ of an
\elpi-experiment, as mentioned before:
\begin{align}
    I_i(t)\ \leftrightarrow\ \dstokesi(\rhopi)\,,
    \qquad i\in\{1,\dots,14\}\,.
\end{align}
The quantities $\dstokesi$ and $\rhopi$ denote the 
geometric mean of the Stokes equivalent diameters
and the particle density associated with
\elpi-stage~$i$, respectively.
In Sec.\,\ref{sec:results}, these combinations
of measured electrical currents and deduced
diameters are used to compute extinction
coefficients of different aerosols.\\

\subsubsection*{Aerosol density $\rhoas$}

An important quantity the total amount of light
extinction depends on is the overall mass density
$\rhoas$ of the aerosol.
One advantage of \elpi-measurements in general is 
that the quantity can be deduced from obtained
experimental data.
This is done by summing up particle masses measured by
the different \elpi-stages and normalizing to unit volume,
as done e.\,g.\ in Eq.\,\eqref{eq:elpi_densities_b}.
Considering the first $n$ \elpi-stages, for example,
the density of the corresponding aerosol part is given by
\begin{align}
    \rhoas^{(n)}(t)\defeq\sum_{i=1}^{n}\rho_{\text{M},i}(t)\,.
    \qquad n\in\{1,\dots,14\}\,,
    \label{eq:rho_soot}
\end{align}
In the considered setup, each $\rhoas^{(n)}$ is
mainly governed by the particle densities $\rhopi$.
This dependence, however, is only weak.
The reason is that individual mass concentrations
$\rho_{\text{M},i}$ are determined indirectly from
volumes and mass densities of the Stokes equivalent
particles, see Eq.\,\eqref{eq:elpi_densities_b}.
Since an increase of the particle density 
at the same time corresponds to smaller Stokes equivalent
diameters and therefore to smaller volumes, both effects
compensate each other and result
in a weak dependence on the actual value of $\rhopi$.
In other words, obtained values for $\rhoas^{(n)}$ are
approximately a direct output of an \elpi-measurement
as a consequence of internal conversion processes.%
\footnote{
    Overall, \elpi conversion factors for mass
    density are proportional to
    $\rho_M\!\propto\!\rhopi\!\cdot\!(\dstokes)^{3-a}$
    with $a\in\{1.225,1.515,1.085\}$, depending on
    the respective Stokes equivalent diameter, see
    Eq.\,\eqref{eq:elpi_densities_b} and
    Eqs.\,\eqref{eq:conversion_factor} in Appendix~A1.
    Stokes equivalent diameters in turn are 
    proportional to
    $\dstokes\!\propto\!(\rhopi)^{-\frac{1}{2}}$
    which can be seen in Eq.\,\eqref{eq:ratio}.
    In total, conversion factors are therefore
    proportional to
    $\rho_M\!\propto\!(\rhopi)^{\frac{a-1}{2}}$
    with a maximum dependence given by
    $\rho_M\!\propto\!(\rhopi)^{\frac{1.515-1}{2}}
    \!=\!(\rhopi)^{0.2575}$.
    Doubling $\rhopi$, for example, then
    only leads to a mass concentration increase
    of at most~$20\,\%$.
    In practice, the increase is typically much
    smaller in the order of a few percent,
    which is shown in Sec.\,\ref{sec:results}.
    }

\subsubsection*{Average distance between aerosol particles}

A quantity closely related to the aerosol density is
the average distance $\overline{l}_{d}$ between the
aerosol particles.
It can be obtained via dividing the unit volume $V_{\rm u}$
by the absolute particle number~$N_{\rm u}$ present
in that volume,
\begin{align}
    \overline{l}_{d}^3=\frac{V_{\rm u}}{N_{\rm u}}
        =\frac{\qty{1}{\cm\cubed}}{N_{\rm u}}\,.
\end{align}
As it turns out, in the experiments conducted in this
contribution, the maximum particle number within the steady
states is $N_{\rm u}\approx5\!\cdot\!10^6$, resulting
in a minimal distance of
$\overline{l}_{d}\!\approx\!\qty{60}{\text{\textmu}\m}$.
Accordingly, the average distance between the 
aerosol particles is much larger than the size of
the equivalent spheres which is (much) smaller than
$\dstokes\!=\!\qty{10}{\text{\textmu}\m}$.
Aerosol particles can therefore be assumed to interact
independently of each other in terms of
single-particle interactions \cite{bohren2009}.\\


\subsection{Transition measurements with MIREX}
\label{sec:visibility_measurements}

As a second, independent measuring system for light interaction
of aerosol particles we use the well-established extinction
measuring equipment MIREX \cite{mirex}.
The device analyses smoke densities based on optical measurements
in the infrared domain of the spectrum.
More precisely, the system uses pulsed light from
a diode as light source and sends it to a reflecting mirror
in approximately one meter distance from the emitter.
The reflected light is then measured by a receiver, resulting in
an overall length of the optical path of $l\!=\!\qty{2}{\m}$.
Accordingly, the MIREX constantly measures light-intensity
reduction in an area of one meter extension.
The specification of the light spectrum is such that
the main part of the power is emitted in the wave length
range between $\qty{0.8}{\text{\textmu} m}$ and
$\qty{0.95}{\text{\textmu} m}$.
The peak wave length of the distribution is
$\lin\!\approx\!\qty{.88}{\text{\textmu} m}$.

\begin{figure}[t!]
    \centering
    \includegraphics[width=0.95\linewidth]{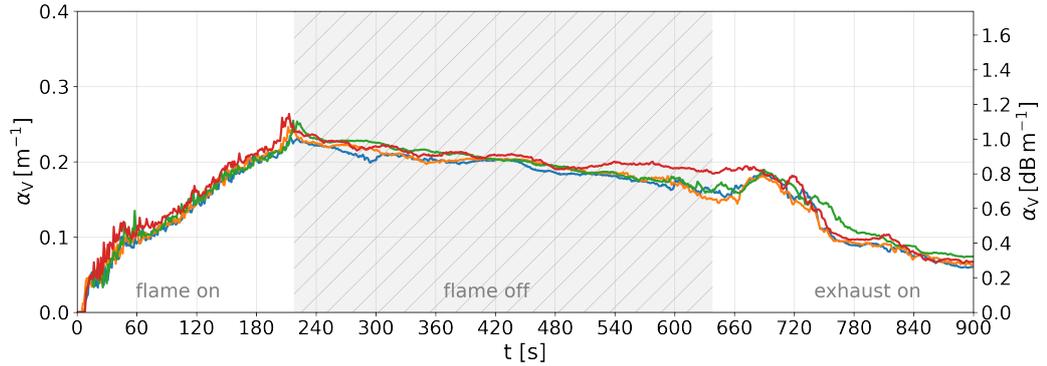}
    \caption{
        \textbf{Time-dependent light extinction measured by
        the MIREX-system.}
        Shown are results for soot from flaming
        $n$-heptane fires obtained in four independent
        experiments. Details of the experimental setup are 
        provided in Sec.\,\ref{sec:results}.
        \label{fig:tf5_all}}
\end{figure}

As an output of an individual MIREX-measurement, one obtains
time-dependent values for the (volume) extinction coefficient
$\alpha_{\rm V}$ that enters the Beer–Lambert–Bouguer law
\cite{bohren2009}
\begin{align}
    \frac{I(t)}{I_0}=e^{-\alpha_{\rm V}(t)\cdot l}\,.
    \label{eq:beer_lambert_bouguer}
\end{align}
Here, $I(t)$ is the measured time-dependent light intensity
and $I_0$ is the intensity of the light source without aerosol.
Sample output of the conducted experiments is shown in
Fig.\,\ref{fig:tf5_all}.


\newpage
\subsection{Light extinction in classical electrodynamics}
\label{sec:theoretical_framework}

\begin{figure}[t!]
    \centering
    \begin{subfigure}{.49\textwidth}
      \caption{$m\!=\!1.5\!+\!0.0\,i$}
      \vspace{-17pt}
      \includegraphics[width=1.0\linewidth]{%
        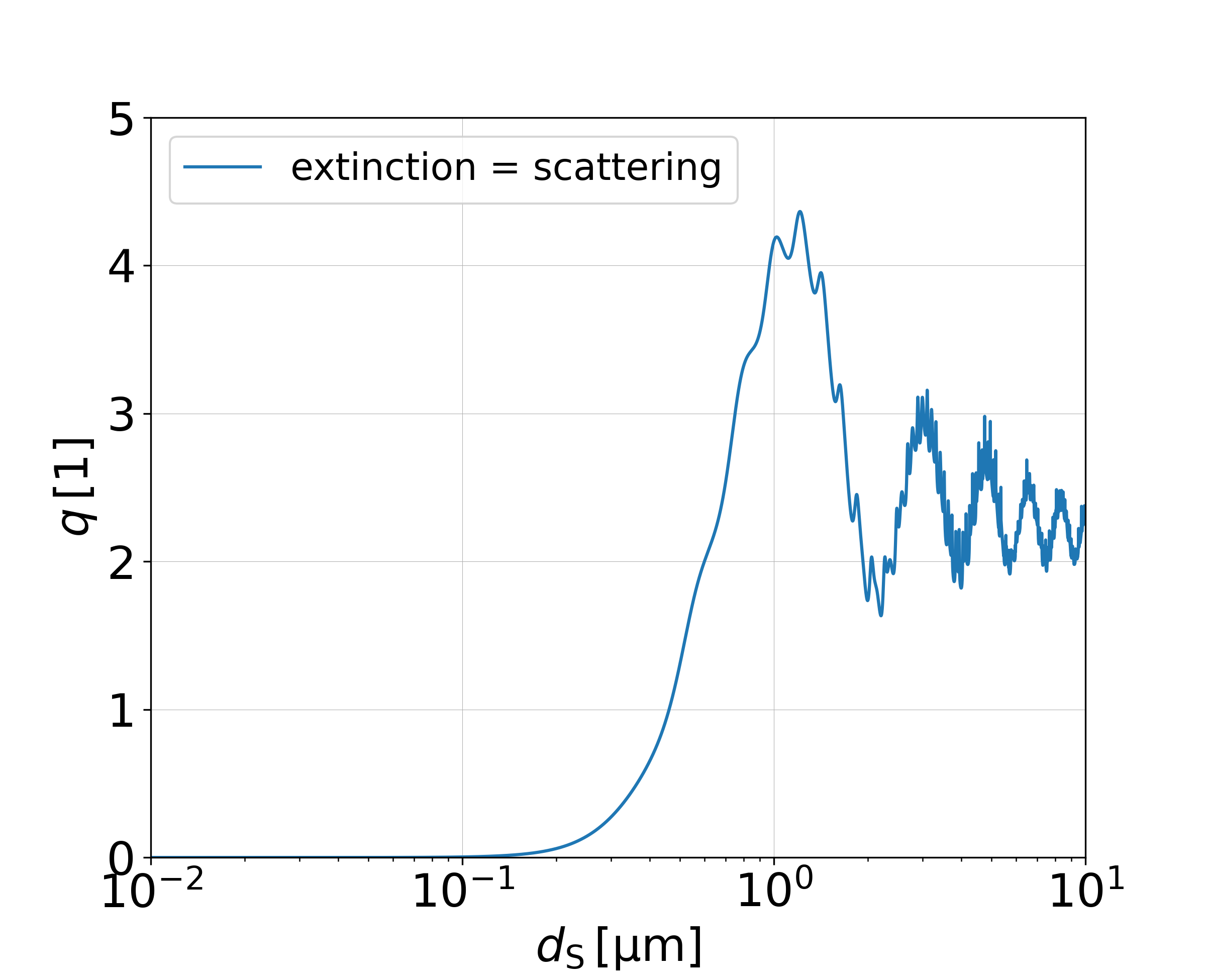}
    \end{subfigure}
    \begin{subfigure}{.49\textwidth}
        \caption{$m\!=\!1.5\!+\!0.5\,i$}
        \vspace{-17pt}
        \includegraphics[width=1.0\linewidth]%
            {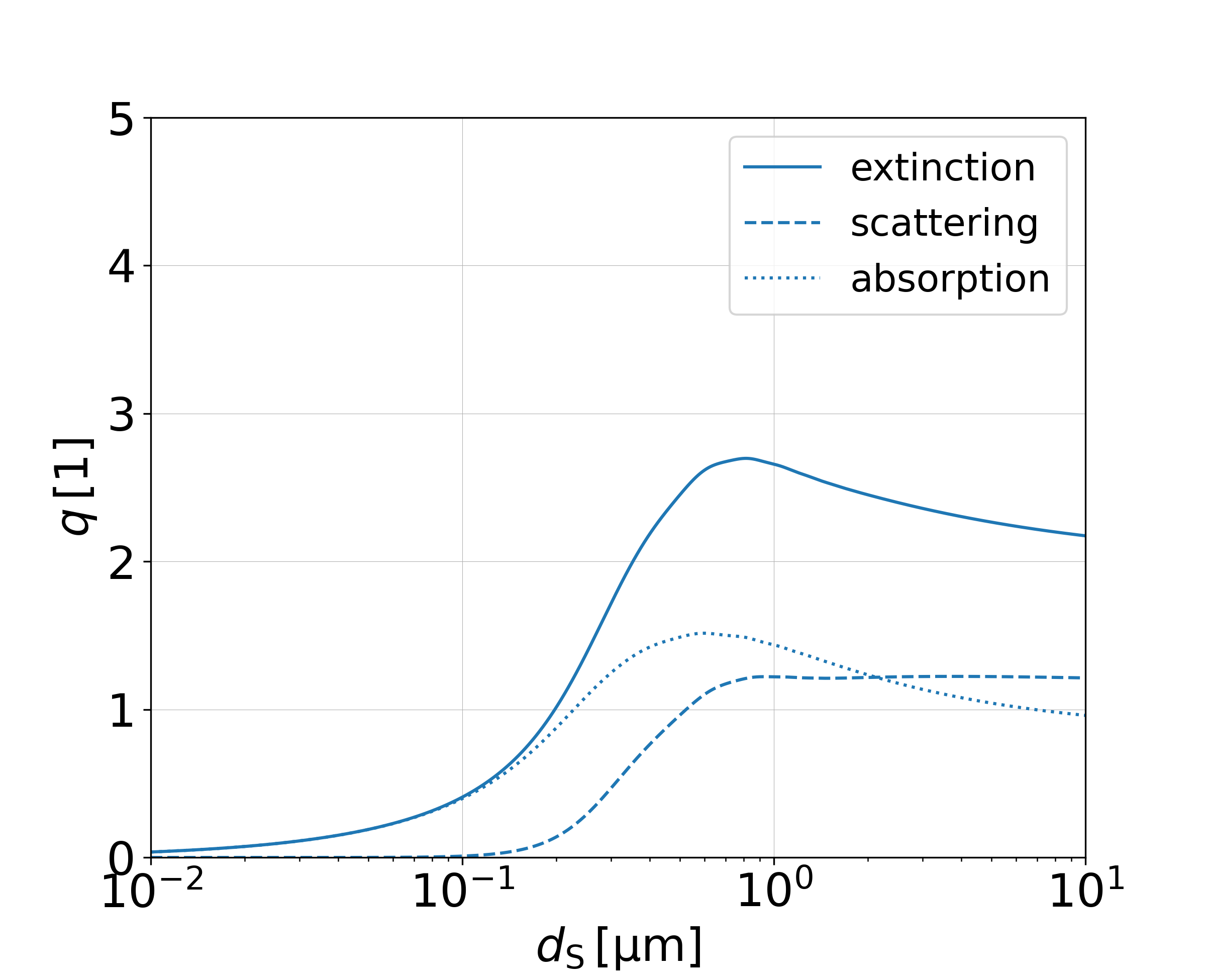}
    \end{subfigure}
    \caption{
        \textbf{Extinction efficiency depending on the equivalent
        diameter and refractive index of the interacting particle.}
        Shown are results for a non-absorbing aerosol (left)
        and an absorbing one (right).
        The wave length of the incident light is
        $\lin\!=\!\qty{0.88}{\text{\textmu}\m}$ in both cases.
        The shown diameter range corresponds to the one
        covered by the \elpi-device.
        \label{fig:extinction_efficiency}}
\end{figure}

One main goal of the present contribution is to
describe light extinction in the presence of
different aerosols by using \elpi-measurements
(discussed in Sec.\,\ref{sec:soot_measurements})
and MIREX measurements (discussed in Sec.\,%
\ref{sec:visibility_measurements}).
In the following, we provide theoretical
background that is necessary to evaluate
\elpi-data accordingly.
Our approach is the following:
In a first step, we start with a basic theory
that is as simple as possible but covers,
nevertheless, all relevant aspects of the
underlying physical processes.
Then, we discuss possible modifications
of the theory that may (or may not) improve the
predictive power of the initial model.

In a first-order approach, the interaction of a 
light beam with a spatially extended aerosol is an
electrodynamic process that can be adequately described
by classical electrodynamics where relativistic and
quantum effects are neglected.
Moreover, to obtain meaningful results in the context
of the present work, we make the following additional
assumptions:
\begin{itemize}
    \item In Sec.\,\ref{sec:soot_measurements} it is shown
        that the \elpi-device assigns dedicated Stokes
        equivalent diameters~$\dstokes$ to each of the
        $14$ measuring channels.
        In the following, we therefore assume interacting
        particles to be \textbf{homogeneous spheres},
        each with an individual geometrical
        diameter $\dgeo\!\equiv\!\dstokes$ and refractive index $m$.
        All particles are monodisperse and interact
        independently of each other in terms of single-particle
        interactions.
    \item In Sec.\,\ref{sec:visibility_measurements}
        it is mentioned that the intensity distribution
        of the MIREX light source is comparatively narrow.
        In the following, we therefore assume the
        incident light beam to consist of      
        \textbf{plane, unpolarized}, and
        \textbf{monochromatic electromagnetic waves}.
        As wave length~$\lin$ we take the peak
        wave length of the intensity distribution
        which is in the infrared 
        and given by $\lin\!=\!\qty{0.88}{\text{\textmu}\m}$.
\end{itemize}
Of course, these assumptions only represent a first-order
description of the optical effects actually taking place.
We show, however, that even with these strong
restrictions it is possible to obtain reliable numerical
results under various different conditions.
In order to show this, we explicitly determine the
following quantities:\\

\subsubsection*{Extinction efficiency  $q_{\rm ext}$}

This parameter quantifies the ability of a single particle to
attenuate parts of a light beam from its original path.
Starting from Maxwell's equations, it can be
computed from basic principles and contains
the two major components \textit{scattering} and
\textit{absorption} \cite{bohren2009}:
\begin{align}
    q_{\rm ext}
    =q_{\rm sca}+q_{\rm abs}
    =q_{\rm ext}(\lin,\dstokes,m)\,.
    \label{eq:qext}
\end{align}
Absorption is not present for every aerosol and governed
by the imaginary part of the refractive index $m$.
Moreover, extinction efficiency depends on the
wave length $\lin$ of the incident light as well as on the
equivalent diameter $\dstokes$ of the interacting particle.
It can be computed e.\,g.\ by using the freely available
software \textit{MiePlot} \cite{mieplot} which itself
is based on the code published in the appendix of
\cite{bohren2009}.
Sample output of the software for two different 
refractive indices is shown in
Fig.\,\ref{fig:extinction_efficiency}.
As can be seen, the measuring range of the
\elpi-device, which ranges from $\qty{0.006}{\text{\textmu}\m}$
to $\qty{10}{\text{\textmu}\m}$ \cite{elpi},
covers the global maximum of the efficiency
distributions, respectively.\\

\subsubsection*{Extinction cross section $\sigma_{\rm ext}$}

\begin{figure}[t!]
\centering
\begin{subfigure}{.49\textwidth}
    \includegraphics[width=1.0\linewidth]{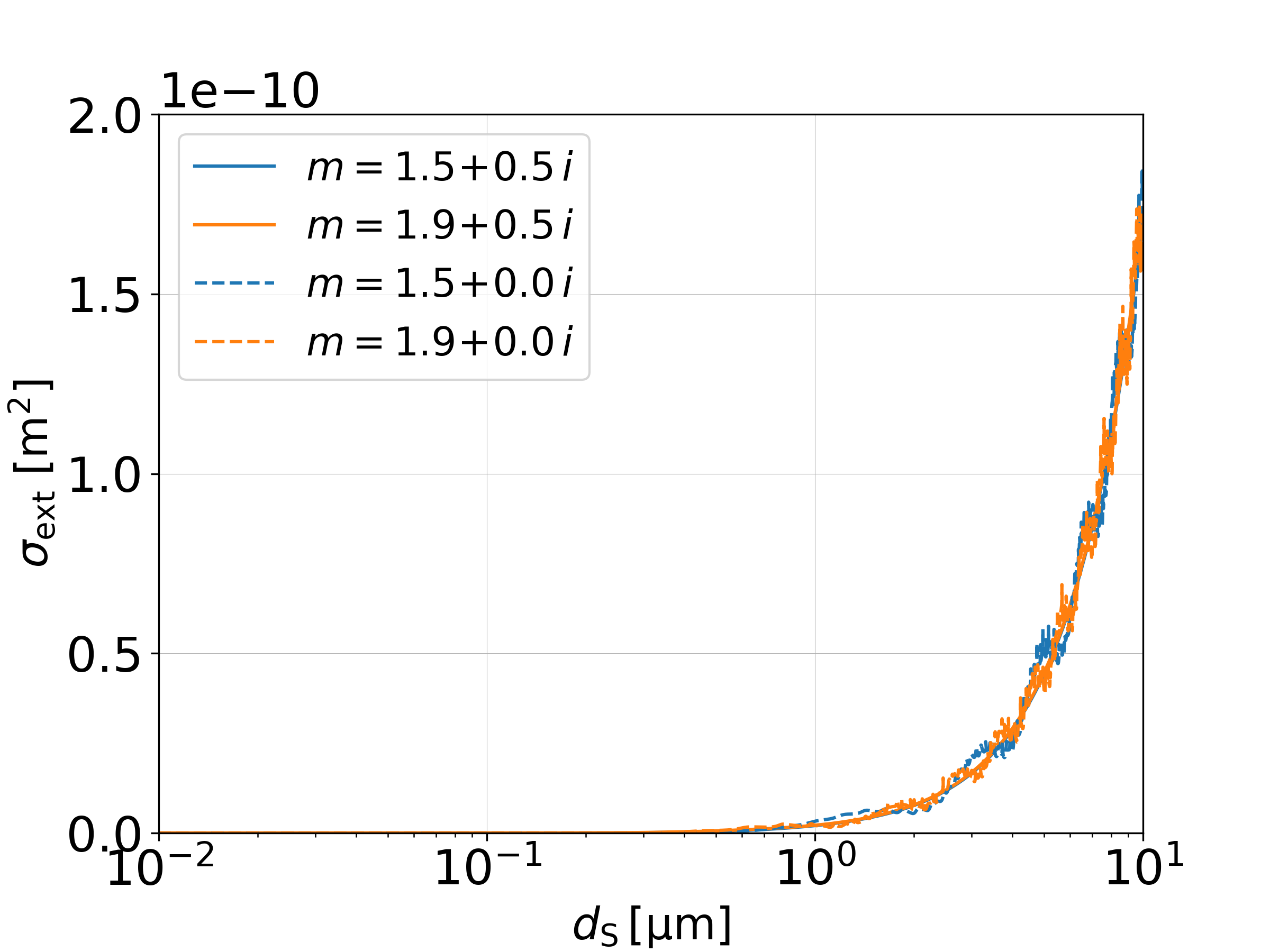}
\end{subfigure}
\begin{subfigure}{.49\textwidth}
    \includegraphics[width=1.0\linewidth]{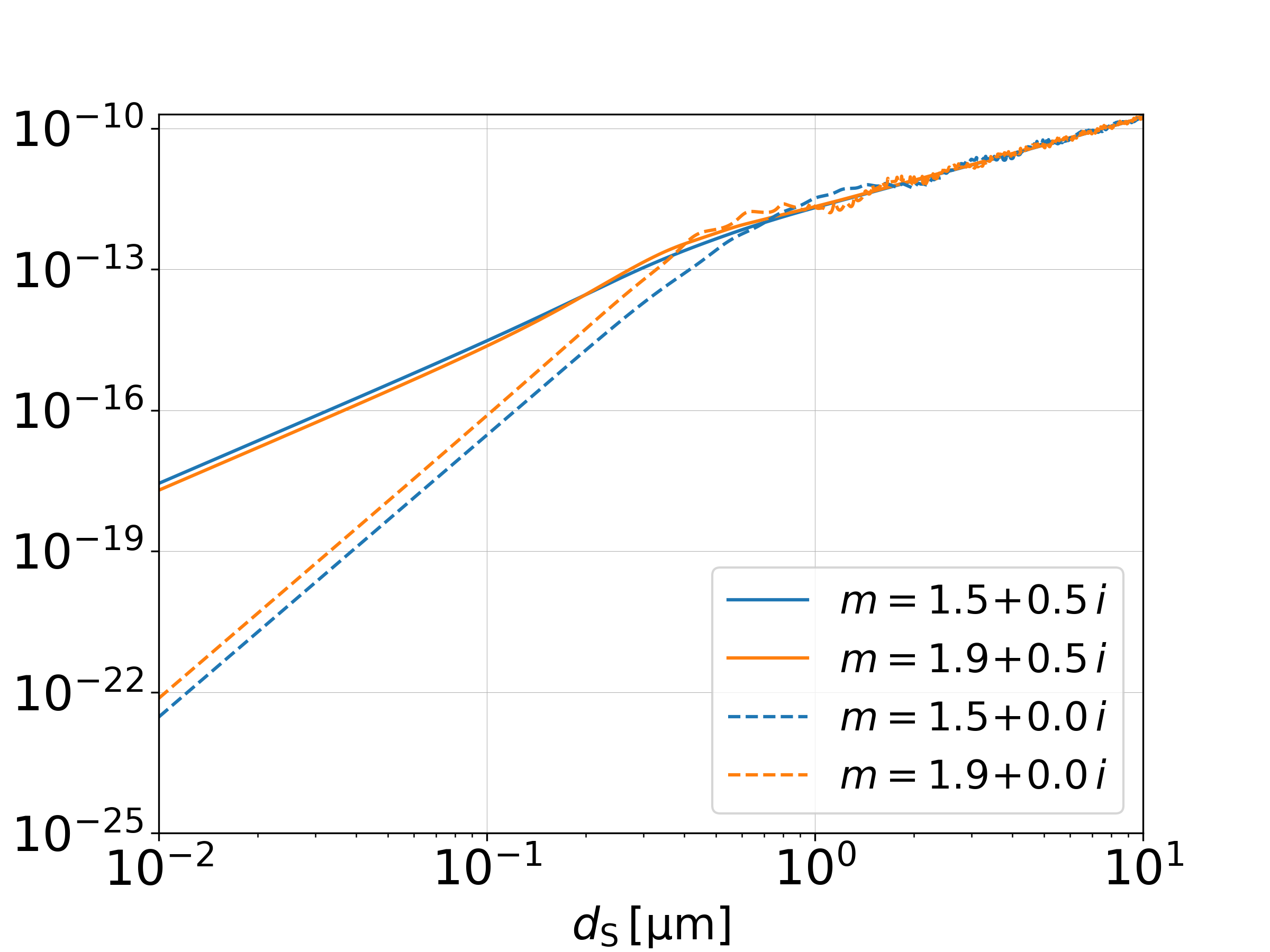}
\end{subfigure}
\caption{
    \textbf{Extinction cross section of one interacting
    particle depending on the equivalent diameter of the particle.}
    Shown are results obtained from Eq.\,\eqref{eq:sigma}
    for different refractive indices.
    Both diagrams have in fact the same content.
    In the right one, however, the $y$-axis has a
    logarithmic scale.
    The wave length of the incident light is
    again $\lin\!=\!\qty{0.88}{\text{\textmu}\m}$.
    \label{fig:extinction_cross_section}}
\end{figure}

This quantity is directly related to the extinction
efficiency.
In contrast to the latter, however, it is a
physical observable that can actually be measured
experimentally. 
Cross section and efficiency are related through the
\textit{visible} particle area in the direction of
the light beam which is a circle in case spherical
objects are considered.
The extinction cross section and its constituting
parts are for an equivalent particle therefore given by
\begin{align}
    \sigma_{j}(\lin,\dstokes,m)
    \defeq q_{j}(\lin,\dstokes,m)\cdot\pi\cdot
    \Big(\frac{\dstokes}{2}\Big)^2 ,
    \qquad j\in\{\rm ext,\,sca,\,abs\}\,.
    \label{eq:sigma}
\end{align}
Examples of extinction cross sections
for different refractive indices are shown in
Fig.\,\ref{fig:extinction_cross_section}.
In first approximation, the distributions are
monotonously increasing.
Larger particles therefore have a larger
extinction cross section than smaller ones.
 Moreover, relevant dependence on the refractive
index is only present for equivalent diameters
smaller than the wave length of the incident light.
Particles larger than that almost behave
identically which corresponds to the principles
of geometrical optics
\cite{bohren2009}.

\subsubsection*{Extinction coefficient%
\footnote{In some references, this quantity is also denoted as
    `attenuation' coefficient, see e.\,g.\ \cite{bohren2009}.}
$\alpha_{\rm V}$}

This quantity is the cumulative sum of all particle
extinction cross sections of an aerosol.
Several different definitions can be found 
in the literature.
In the present contribution, we use the so-called
\textit{volume extinction coefficient}
which is normalized to unit volume
\mbox{$V_{\rm u}\!=\!\qty{1}{\cm\cubed}$}.
In this form, it is exactly the quantity that enters
the Beer–Lambert–Bouguer law in
Eq.\,\eqref{eq:beer_lambert_bouguer}
\cite{bohren2009}.
Using particle number densities of the $14$ \elpi-channels
defined in Eq.\,\eqref{eq:elpi_densities_a} as well as
extinction cross sections given in Eq.\,\eqref{eq:sigma},
$\alpha_{\rm V}$ can be written as
\begin{subequations}
\label{eq:alpha_V}
\begin{align}
    \alpha_{\rm V}(t)
    &=\sum_{i=1}^{14}
    \alpha_{\text{V},i}(t)
    =\sum_{i=1}^{14}
    \frac{N_i(t)}{V_{\rm u}}\cdot\sigma_{\rm ext}(\dstokesi,m)
    =\sum_{i=1}^{14}
    \rho_{\text{N},i}(t)\cdot\sigma_{\rm ext}(\dstokesi,m)
    \label{eq:alpha_Va}
    \\
    &\stackrel{\eqref{eq:elpi_densities_a},\eqref{eq:sigma}}{=}\pi
   \sum_{i=1}^{14}
       q_{\rm ext}(\dstokesi,m)
       \cdot\left(\frac{\dstokesi}{2}\right)^2
       \cdot X_{N,i}(\dstokesi)
       \cdot I_{i}(t)\,.
   \label{eq:alpha_Vb}
\end{align}
\end{subequations}
Eq.\,\eqref{eq:alpha_Vb} is one of the most important
formulas in the present contribution.
In this formula, extinction coefficients are expressed
in terms of extinction efficiencies $q_{\rm ext}$
(and therefore depend on the refractive index $m$),
equivalent diameters $\dstokesi$
(and therefore depend indirectly on corresponding
particle densities $\rhopi$, see Eq.\,\eqref{eq:ratio}),
conversion factors $X_{N,i}$,
and electrical currents~$I_{i}(t)$
measured by the \elpi-device.

In the following, we assume the same \textit{effective}
particle density $\rhop$ for each \elpi-channel,
\begin{align}
    \rhopi\equiv\rhop\,,
    \qquad i\in\{1,\dots,14\}\,,
    \label{eq:rhop}
\end{align}
which turns out to be sufficient to consistently
combine model predictions and experimental results
from MIREX- and \elpi-measurements.%
\footnote{
    The authors of the present contribution are
    aware of the fact that individual particle
    densities $\rhopi$ can actually be assumed to
    be different for individual \elpi-channels,
    see e.\,g.\ \cite{COUDRAY2009}.
    We further comment on limitations of
    assumption~\eqref{eq:rhop} in the course
    of the contribution, in particular in
    Sec.\,\ref{sec:results}.}
Using Eq.\,\eqref{eq:rhop}, we therefore only have
\textit{three independent aerosol properties} in
Eq.\,\eqref{eq:alpha_Vb} that can be used to adjust
the model, namely
\begin{itemize}
        \item the real part $\mathfrak{Re}(m)$
        of the refractive index\\
        which describes \textit{scattering} properties
        of the aerosol particles,
    \item the imaginary part $\mathfrak{Im}(m)$
        of the refractive index\\
        which describes \textit{absorbing} properties
        of the aerosol particles, and
    \item the effective particle density $\rhop$.
\end{itemize}
In Sec.\,\ref{sec:results}, we consider different aerosols
and show that, using known literature values of
$m$ and $\rhop$,
it is possible to consistently describe all obtained
experimental results within a unified framework.
For some applications it turns out to be useful
to also define the cumulative extinction coefficient
obtained from the first $n$ \elpi-channels only,
similar to Eq.\,\eqref{eq:rho_soot}:
\begin{align}
    \alpha_{\rm V}^{(n)}(t)
        \defeq \sum_{i=1}^{n}\alpha_{\text{V},i}(t)\,,
        \qquad n\in\{1,\dots,14\}\,.
    \label{eq:alpha_Vn}
\end{align}
\vspace{5pt}

\subsubsection*{Mass-specific extinction coefficient $\alpha_{\rm M}$}

\begin{figure}[t!]
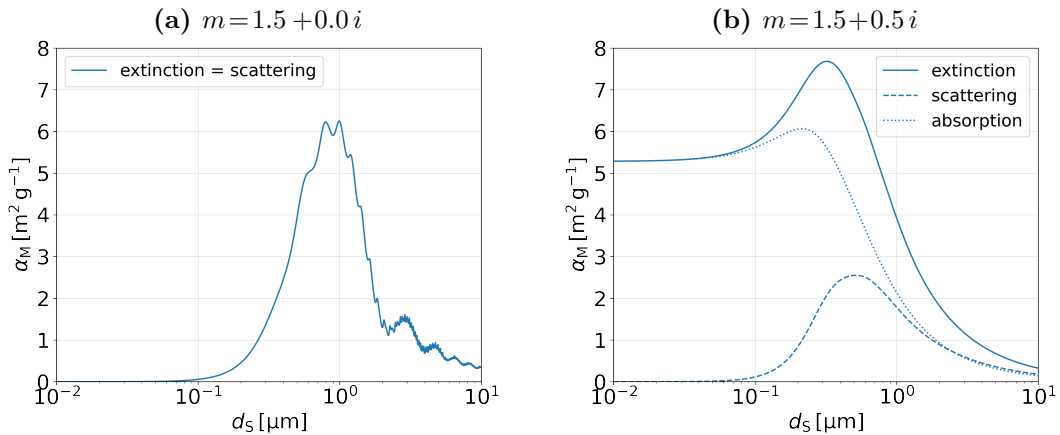

    \centering
    \begin{subfigure}{0.45\textwidth}
        \caption{
            $m\!=\!1.5+\!0.0\,i$
            \label{fig:alphaMi1a}}
        \vspace*{-17pt}
        \includegraphics[width=\linewidth]{figures/%
             03_alphaMi1_m=1.50+0.00i.png}
    \end{subfigure}
    \begin{subfigure}{0.45\textwidth}
        \caption{
            $m\!=\!1.5\!+\!0.5\,i$
            \label{fig:alphaMi1b}}
        \vspace*{-17pt}
        \includegraphics[width=\linewidth]{figures/%
            03_alphaMi1_m=1.50+0.50i.png}
    \end{subfigure}
    \caption{
        \textbf{Mass-specific extinction of a single aerosol
        particle depending on the equivalent diameter
        of the particle.}
        The curves are obtained from
        Eq.\,\eqref{eq:alpha_Mi} using
        $m\!=\!1.5\!+\!0.0\,i$ (left) and
        $m\!=\!1.5\!+\!0.5\,i$ (right).
        The particle density and the wave length of
        the incident light are in both cases fixed as
        $\rhop\!=\!\qty{1}{\g\per\cm\cubed}$ and
        $\lin\!=\!\qty{0.88}{\text{\textmu}\m}$,
        respectively.
        \label{fig:alphaMi1}}
\end{figure}

This quantity is closely related to the extinction coefficient
$\alpha_{\rm V}$ and describes light extinction per
mass density unit of the aerosol.
According to \cite{bohren2009}, it is defined as 
\begin{align}
    \alpha_{\rm M}
    &\defeq \frac{\alpha_{\rm V}(t)}{{\rho}_{\rm aerosol}(t)}\,.
    \label{eq:alpha_M}
\end{align}
To be able to estimate the size of $\alpha_{\rm M}$, we first 
consider a single equivalent particle and determine its
contribution to the total value of $\alpha_{\rm M}$.
Using Eqs.\,\eqref{eq:alpha_Vb}, \eqref{eq:rho_soot}, and
\eqref{eq:elpi_densities}, mass-specific extinction
of a single particle associated with \elpi-channel~$i$
can be written as
\begin{align}
    \alpha_{\text{M},i}
    \defeq\frac{\alpha_{\text{V},i}}{\rho_{\text{aerosol},i}}
    =\frac{3}{2}%
        \frac{q_{\rm ext}(\dstokesi,m)}{\dstokesi\cdot\rhop}\,.
    \label{eq:alpha_Mi}
\end{align}
Accordingly, $\alpha_{\text{M},i}$ is a constant in time
that depends on the extinction efficiency,
on the corresponding Stokes equivalent
diameter (directly and indirectly),
as well as on the particle density.
Exemplary results for two different refractive indices
are shown in Fig.\,\ref{fig:alphaMi1}.
Since $q_{\rm ext}\!\to\!2$ in the limit of
large (equivalent) particles, see
Fig.\,\eqref{fig:extinction_efficiency},
mass-specific extinction vanishes 
for large $\dstokes$:
\begin{align}
    \lim_{\dstokesi\,\to\,\infty}\alpha_{\text{M},i}=0\,.
    \label{eq:alphaM_limit}
\end{align}
The reason for this effect is that $\alpha_{\rm V}$ is
related to the circular-shaped \textit{surface}
of the equivalent particle in the direction
of the light beam (and therefore $\propto\!\dstokes^2$,
see Eq.\,\eqref{eq:sigma}),
whereas $\rho_{\text{aerosol}}$ is related to the
particle \textit{volume} (and therefore
$\propto\!\dstokes^3$, see Eq.\,\eqref{eq:elpi_densities_b}).
For increasing values of $\dstokes$, the overall
dependence is therefore $\propto\!\dstokes^{-1}$.
In other words, contributions to mass-specific
extinction from the largest particles measured
by the \elpi-device are smaller than the ones
from intermediate-size particles
since larger particles proportionally contribute
more to the aerosol density than to
light extinction.
Accordingly, $\alpha_{\rm M}$ in
Eq.\,\eqref{eq:alpha_M} describing the whole
aerosol does not depend on numerical values
of $\alpha_{\rm V}$ and $\rhoas$ alone, but also
on the \textit{size-distribution} of the individual
equivalent diameters.
This fact is a general prediction of classical
electrodynamics and already evident within the
measuring range of the \elpi-device, as can be seen
in Fig.\,\ref{fig:alphaMi1}.


\newpage
\section{Application of the model}
\label{sec:results}

In the previous section, we deduce Eq.\,\eqref{eq:alpha_Vb}
in which light-intensity reduction is expressed
in terms of the parameters $m$ and $\rhop$.
For a given aerosol, it is in principle possible
to use known literature values of the input parameters
to obtain corresponding results for extinction coefficients.
This, however, is complicated by the fact 
that both quantities are comparatively
difficult to determine experimentally and, more
importantly, through distinct independent measurements.
Typically, the literature does therefore not provide
combinations of $m$ and $\rhop$
but only values or value ranges
for one parameter at a time.
Then, it is often not clear if considered measurements
have actually been conducted under overall comparable
experimental conditions.

In this contribution, we therefore follow a
different approach.
Since light extinction is determined independently
through MIREX-measurements in each of
the conducted experiments, we consider $\alpha_{\rm V}(t)$
and its time average $\overline{\alpha}_{\rm V}$
as known and try to reproduce obtained MIREX-results
by finding appropriate combinations of $m$ and $\rhop$
in Eq.\,\eqref{eq:alpha_Vb}.
We do this by scanning the refractive index
over the known range of literature values.
For each fixed $m$, we then find the unique numerical
value of $\rhop$ for which the time average of
$\alpha_{\rm V}(t)$ is the same for both measurements,
i.\,e.\ $\overline{\alpha}_{V,\text{\elpi}}\!\equiv\!%
\overline{\alpha}_{V,\text{MIREX}}$.
Although in principle all \elpi-channels can be used
in the evaluation of the measurement data,
in the following we only consider the first $13$~stages.
This is done since contributions of channel~%
$14$ to total light extinction are only small
(less than $\qty{1}{\percent}$ in case of paraffin
and less than $\qty{5}{\percent}$ in case of TF5).
In this way, however, the total amount and therefore the
quality of valid data that can be evaluated increases
significantly. In this setup, the measuring range of the
\elpi is approximately given by
$\qty{0.006}{\text{\textmu}\m}\!<\!\dstokes\!<\!%
\qty{4}{\text{\textmu}\m}$.


\subsection{Extinction coefficient \texorpdfstring{$\alpha_{\rm V}$}{alphaV}}
\label{sec:ext_coeff}

\subsubsection*{Paraffin  aerosol}

\addtocounter{footnote}{+1}
\footnotetext{
    Also smaller values of $\mathfrak{Re}(m)$
    can be found in the literature,
    e.\,g.\ in Refs.\,\cite{Ma2022,Zhao2018}.
    These values, however, have been obtained for
    larger ambient temperatures of
    $T_{\rm exp}\!\geq\!\qty{40}{\degree C}$.}
\addtocounter{footnote}{-1}

All paraffin experiments discussed in this contribution
have been conducted at the department of
Communication Systems at the University of Duisburg-Essen
using the test duct described in~\cite{Muller:2023}.
In each of the experiments, a `ramp' setup is 
considered in which the aerosol concentration is
successively increased by external supply before
the supply is stopped and a steady state is forming.
As value ranges of the model parameters we take
into account
\begin{subequations}
    \label{eq:parameters1}
    \begin{align}
        \mathfrak{Re}(m) &= (1.3-1.5)
            \phantom{()\,\unit{\g\per\cm\cubed}}
            \text{\cite{Cooper:82,Li2015,kontges:2016}\,,%
                \footnotemark}
            \label{eq:parameters1a}
        \\[-2.9pt]
        \mathfrak{Im}(m) &= 0
            \phantom{,(-1.5)\quad \unit{\g\per\cm\cubed}()}
            \text{\cite{Cooper:82,Li2015,kontges:2016}}\,,
        \\[-2.9pt]
        \rhop & = (0.8-0.9)\,\unit{\g\per\cm\cubed}
            \text{\ (at $20\, ^{\circ} C$)}\,.
             \label{eq:parameters1c}
    \end{align}
\end{subequations}
Note, that the imaginary part of the
refractive index vanishes.
Accordingly, the aerosol does not absorb
light and extinction is determined exclusively by
scattering processes, see also Eq.\,\eqref{eq:qext}.

\begin{figure}[ht!]
    \centering
    \includegraphics[width=.9\linewidth]{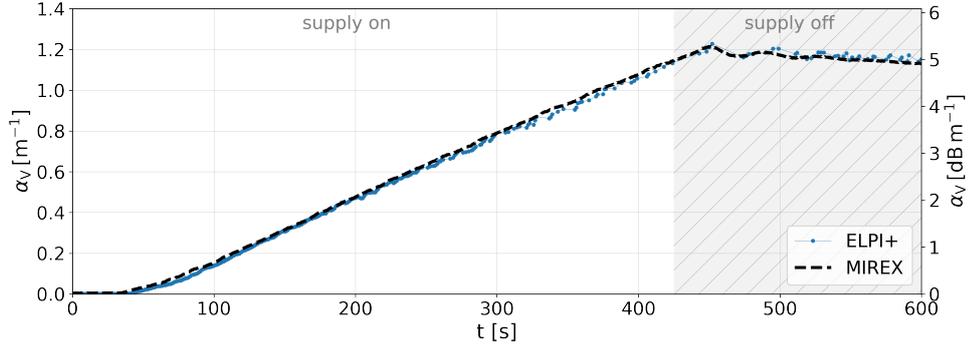}
    \caption{
        \textbf{Time-dependent extinction coefficient obtained
        from a paraffin ramp experiment.}
        The black curve is obtained from the MIREX-system,
        the blue curve from Eq.\,\eqref{eq:alpha_Vb}
        using $m\!=\!1.42$ and $\rhop\!=\!\qty{0.85}{\g\per\cm\cubed}$.
        The white area indicates the time span in which
        the aerosol concentration is increased by external supply.
        In the grey area, the supply is stopped and the system
        passes into a steady state.
        \label{fig:paraffin1}}
\end{figure}

The comparison of MIREX- and \elpi-measurements
for one exemplary paraffin ramp experiment is shown
in Fig.\,\ref{fig:paraffin1}.
As can be seen, almost perfect agreement between
the two measuring systems can be obtained using
appropriate model parameters, in this case
$m\!=\!1.42$ and
$\rhop\!=\!\qty{0.85}{\g\per\cm\cubed}$.
It is important to emphasize that these numerical
values are simultaneously compatible with
the value ranges in Eqs.\,\eqref{eq:parameters1}
and, therefore, with known results from the literature.
This fact serves as a first indication that the
developed model is able to correctly
describe MIREX- and \elpi-measurements
without the need of introducing additional
correction or conversion factors.
Among other things, this comprises the
correct description of different experimental
phases (`ramp' phase, steady state) as well
as subtleties in the time development
of~$\alpha_{\rm V}$
(in Fig.\,\ref{fig:paraffin1} e.\,g.\ for
times $\qty{450}{\s}\!<\!t\!<\!\qty{500}{\s}$).

\begin{figure}[ht!]
    \centering
    \parbox{.45\linewidth}{
    \begin{subfigure}{.45\textwidth}
        \includegraphics[width=\linewidth]{
            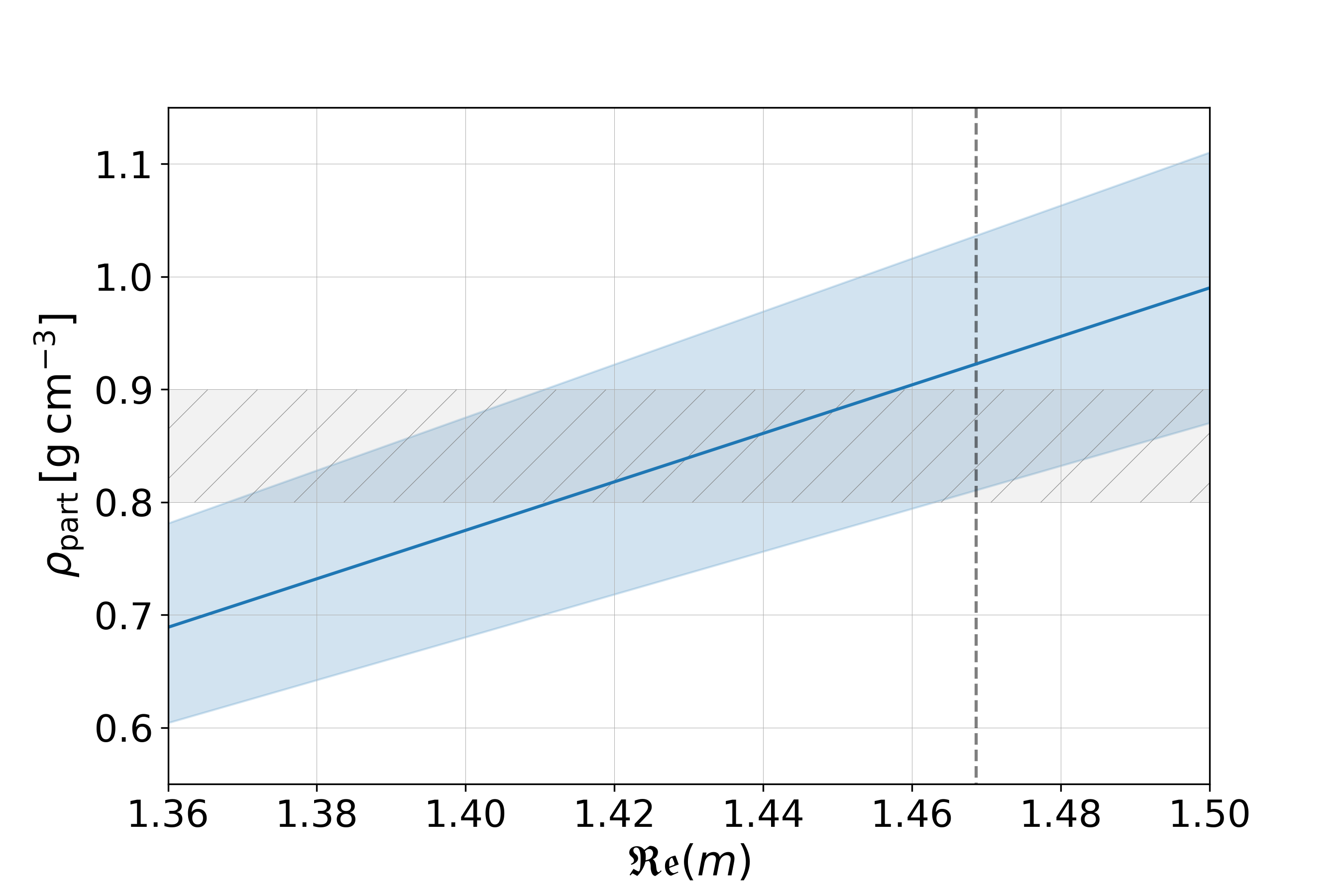}
    \end{subfigure}
    }
    \parbox{.4\linewidth}{
        \begin{tabular}{c c}
                  $\mathfrak{Re}(m)$ &
                  \quad $\rhop\,[\unit{\g\per\cm\cubed}]$ \\
                  \hline
                  $1.30$ & \quad $0.56\pm0.06$ \\
                  $1.35$ & \quad $0.68\pm0.07$ \\
                  $1.40$ & \quad $0.77\pm0.08$ \\
                  $1.45$ & \quad $0.88\pm0.12$ \\
                  $1.50$ & \quad $0.99\pm0.13$ 
        \end{tabular}
    }
    \caption{
        \textbf{Values of $\rhop$ for which MIREX-and
        \elpi-measurements of the paraffin aerosol agree best.}
        Shown is the dependence on the value of $\mathfrak{Re}(m)$
        (blue solid line
        and corresponding values in the table),
        including the $3\sigma$-range
        (blue band and uncertainties in the table).
        The grey band indicates the value range in
        Eq.\,\eqref{eq:parameters1c};
        the dashed vertical line refers to the
        refractive index obtained from an analytical
        formula provided in \cite{Cooper:82}.
        \label{fig:paraffin_parameters}}
\end{figure}

It is important to notice that agreement
between MIREX- and \elpi-measurements is not only
obtained for one particular combination of
$m$ and $\rhop$ but for different ones.
Considering the average of eight
independent paraffin ramp experiments
and scanning over the entire value range in
Eq.\,\eqref{eq:parameters1a}, we obtain further
parameter combinations for which agreement
between MIREX- and \elpi-measurements is achieved.
The results are summarized in Fig.\,%
\ref{fig:paraffin_parameters}.
\vspace*{-20pt}

Since obtained fit values of $\rhop$ differ slightly
from one experiment to another, particle densities
are subject to small variations for which we provide
the $3\sigma$-range.
In general, larger values of $\mathfrak{Re}(m)$
correspond to larger particle densities and
therefore to smaller (Stokes) equivalent particles,
see for instance Eq.\,\eqref{eq:ratio}.
Since smaller particles correspond in first
approximation to less light extinction of the aerosol
(see Fig.\,\ref{fig:extinction_cross_section}),
overall scattering in the presence of a paraffin
aerosol therefore increases with increasing values of
$\mathfrak{Re}(m)$.
In this regard, it should be noted that the
blue band in Fig.\,\ref{fig:paraffin_parameters}
is simultaneously compatible with both value
ranges in Eqs.\,\eqref{eq:parameters1}.

\begin{figure}[t!]
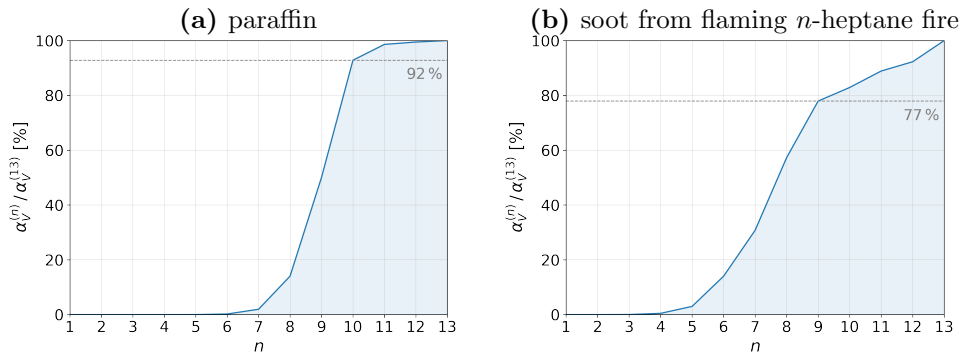

\centering
    \begin{subfigure}{0.40\textwidth}
        \caption{paraffin
            \label{fig:alphaV_cuma}}
        \vspace*{-16pt}
        \includegraphics[width=\linewidth]{figures/%
            paraffin_06_alphaV_cum_13_exp_19.png}
    \end{subfigure}
    \begin{subfigure}{0.40\textwidth}
        \caption{soot from flaming $n$-heptane fire
            \label{fig:alphaV_cumb}}
        \vspace*{-16pt}
        \includegraphics[width=\linewidth]{figures/%
            tf5_06_alphaV_cum_13_exp_4_1.png}
    \end{subfigure}
    \vspace{-10pt}
    \caption{
        \textbf{Relative cumulative contribution of
        the different \elpi-channels to extinction.}
        Shown are results obtained from Eq.\,\eqref{eq:alpha_Vn}.
        In the left diagram, the first $10$ channels
        combined contribute $\qty{92}{\%}$ to entire light
        extinction; the main part ($\qty{91}{\%}$)
        stems from channels $8$, $9$, and~$10$ only.
        In the right diagram, the first $9$ channels
        contribute $\qty{77}{\%}$ to light
        extinction; the main part ($\qty{76}{\%}$)
        stems from channels $6$, $7$, $8$, and~$9$.
        \label{fig:exp_alphaV_cum}}
\end{figure}

Finally, we use Eq.\,\eqref{eq:alpha_Vn}
to evaluate individual contributions of the
different \elpi-channels in the left diagram
of Fig.\,\ref{fig:exp_alphaV_cum}.
As can be seen, the main part of extinction can be
assigned to only a small number of neighboring
\elpi-channels.
\vspace*{-5pt}


\subsubsection*{Soot from flaming $n$-heptane fire}

\begin{figure}[ht!]
\centering
    \includegraphics[width=0.95\linewidth]{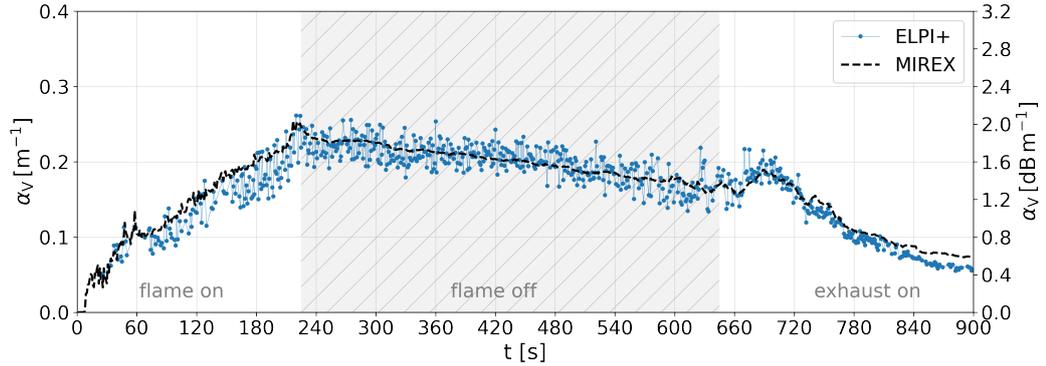}
    \caption{
        \textbf{Extinction coefficient obtained from an $n$-heptane
        fire experiment.}
        The black curve is the result of the
        MIREX-system; the blue curve is the result of 
        Eq.\,\eqref{eq:alpha_Vb} using $m\!=\!1.57\!+\!0.56\,i$
        and $\rhop\!=\!\qty{2.21}{\g\per\cm\cubed}$.
        The differently shaded areas indicate different phases
        of the soot generation, namely the flame-on,
        the flame-off, and the exhaust-on phase, respectively.
        \label{fig:exp_alphaV_41}}
\end{figure}

\begin{figure}[ht!]
    \centering
    \vspace*{-3pt}
    \parbox{.5\linewidth}{
        \begin{subfigure}[b]{0.5\textwidth}
            \includegraphics[width=\linewidth]{
                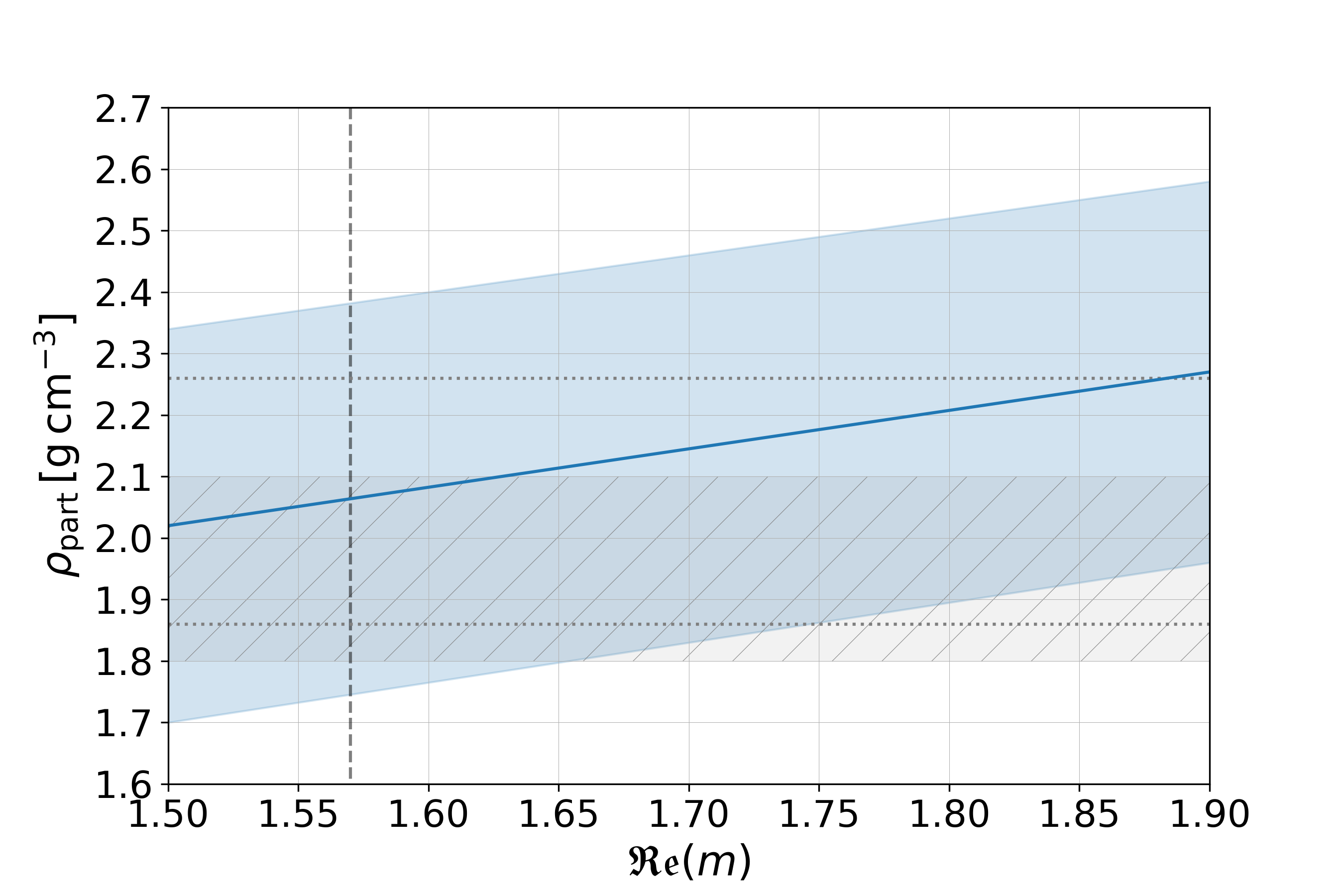}
        \end{subfigure}
        }
    \parbox{.4\linewidth}{
        \centering
        \begin{tabular}{ c c c }
            $\mathfrak{Re}(m)$ &
            $\mathfrak{Im}(m)$ &
            \quad $\rhop\,[\unit{\g\per\cm\cubed}]$\\
            \hline
            $1.50$   & $0.40$ & \quad $1.80\pm 0.28$
            \\
            $1.90$   & $0.40$ & \quad $2.13\pm 0.29$
            \\
            \hline
            $1.50$   & $0.50$ & \quad $2.02\pm 0.32$
            \\
            $1.90$   & $0.50$ & \quad $2.27\pm 0.31$
            \\
            \hline
            $1.57$   & $0.56$ & \quad $2.18\pm 0.35$
            \\ 
            \hline
            $1.50$   & $0.60$ & \quad $2.23\pm 0.36$
            \\
            $1.90$   & $0.60$ & \quad $2.40\pm 0.34$
                        \\
            \hline
            $1.50$   & $0.70$ & \quad $2.45\pm 0.41$
            \\
            $1.90$   & $0.70$ & \quad $2.53\pm 0.37$
        \end{tabular}
        }
    \caption{
        \textbf{
        Values of $\rhop$ for which MIREX-and \elpi-results
        agree best.}
        Shown is the dependence on the values of 
        $\mathfrak{Re}(m)$ and $\mathfrak{Im}(m)$
        (blue solid line and values in the table), 
        including the $3\sigma$-range (blue band and
        uncertainties in the table).
        In the diagram, the imaginary part of $m$
        is fixed as
        $\mathfrak{Im}(m)\!=\!0.5$, whereas
        the dashed vertical line refers to the
        commonly used real part $\mathfrak{Re}(m)\!=\!1.57$ 
        \cite{Smyth_Shaddix_1996}.
        The dotted horizontal lines denote the
        density of crystalline graphite
        ($\rhop\!=\!\qty{2.26}{\g\per\cm\cubed}$)
        and the commonly used density
        for carbonaceous soot
        ($\rhop\!=\!\qty{1.86}{\g\per\cm\cubed}$%
        ~\cite{Dobbins1994ComparisonOA}),
        respectively.
         \label{fig:tf5_parameters_04}}
\end{figure}

A second application case of the model is a room-scale
setup investigating soot from flaming $n$-heptane fires
(test fire TF5 according to standard EN54~\cite{en54}).
Details of the experimental setup are provided in 
\cite{borger:2022}.
In general, the aerosol is characterized
by two major differences compared to paraffin:
First, the particle density $\rhop$ is
\textit{larger}
than unit density $\qty{1}{\g\per\cm\cubed}$.
Stokes equivalent spheres are therefore smaller
than corresponding aerodynamic ones.
Second, soot from an $n$-heptane fire absorbs parts
of a light beam.
Light extinction is therefore a combination of
scattering and absorption processes and not of
scattering alone, see e.\,g.\ Eq.\,\eqref{eq:qext}.
Compared to paraffin, we therefore have three independent
variables instead of two:
\begin{subequations}
\label{eq:parameters}
\begin{align}
    \mathfrak{Re}(m) &= (1.5-1.9)
        \phantom{\,\unit{\g\per\cm\cubed}}
        \text{\cite{%
            Sorensen_2001,
            Smyth_Shaddix_1996,
            Wu1997,
            Chang_1990,
            Dalzell:1969
            }}\,,\\
    \mathfrak{Im}(m) &= (0.4-0.7)
        \phantom{\,\unit{\g\per\cm\cubed}}
        \text{\cite{%
            Sorensen_2001,
            Smyth_Shaddix_1996,
            Wu1997,
            Chang_1990,
            Dalzell:1969
            }}\,,\\
    \rhop & = (1.8-2.1)\,\unit{\g\per\cm\cubed}
        \phantom{}
        \text{\cite{%
            Botero_2018,
            COUDRAY2009,
            Slowik2007,
            Schneider2006,
            Park2004,
            Widmann2003MeasurementOT,
            Hueglin1997,
            Dobbins1994ComparisonOA}}\,.
         \label{eq:parametersc}
\end{align}
\end{subequations}
Choosing the most commonly used refractive index
for carbonaceous soot, $m\!=\!1.57\!+\!0.56\,i$
(see \cite{Smyth_Shaddix_1996} and references therein),
the comparison between MIREX- and ELPI+-measurements
for one exemplary experiment is shown in
Fig.\,\ref{fig:exp_alphaV_41}.
Compared to a paraffin aerosol, the
distribution is characterized by larger
dynamics from one time step to another
which is due to the fact that we are evaluating
a room-scale experiment instead of a bench-scale one.
As before, however, subtleties in the
time development of $\alpha_{\rm V}$ like
different experimental phases
(flame-on, flame-off, exhaust-on phase)
and local extrema (in particular within the
exhaust-on phase) are correctly recovered.
Taking the average of the four independent $n$-heptane
fire experiments in Fig.\,\ref{fig:tf5_all}
and choosing exemplary values for $m$
compatible with Eqs.\,\eqref{eq:parameters},
we again obtain distinct parameter combinations for which
the result of Eq.\,\eqref{eq:alpha_Vb} coincides with
the respective MIREX-measurement, see
Fig.\,\ref{fig:tf5_parameters_04}.
As before, particle densities increase with
increasing values of $\mathfrak{Re}(m)$.
However, the increase is less pronounced in general
and smaller for larger values of $\mathfrak{Im}(m)$.
Overall, it turns out that Eq.\,\eqref{eq:alpha_Vb}
yields reliable results also for flaming
$n$-heptane fires in a room-scale setup.
Thereby, compatible combinations of $m$ and $\rhop$
are in accordance with value ranges 
provided by the literature.

Finally, we again use Eq.\,\eqref{eq:alpha_Vn}
to evaluate individual contributions of the
ELPI+-channels, see the r.\,h.\,s.\ of
Fig.\,\ref{fig:exp_alphaV_cum}.
Compared to paraffin, the slope
of the distribution is less steep.
However, the main part of extinction again stems from
only a small number of neighboring channels.


\subsection{Mass-specific extinction coefficient \texorpdfstring{$\alpha_{\rm M}$}{alphaM}}
\label{sec:specific_extinction}

In the literature, mass-specific extinction has so far
has been studied empirically and mainly in small-scale
experiments under well-controlled and well-ventilated
conditions. A review of such results is provided e.\,g.\ 
in \cite{Widmann_2003}, finding
an overall fit of the collected data as
$\alpha_{\rm M}
\!=\!\qty{4.8081}{\m\squared\per\g}
(\lin/\unit{\text{\textmu}\m})^{-1.0088}$, depending
on the wave length of the incident light.
Transferred to the present MIREX-experiments
($\lin\!=\!\qty{0.88}{\text{\textmu}\m}$),
the fit yields
    \begin{align}
        \alpha_{\rm M}=(5.5\pm 0.6)\,\unit{\m\squared\per\g}
            \qquad(95\%\text{ c.l.})
            \text{\cite{Widmann_2003}}
            \,,
        \label{eq:av_lita}
    \end{align}
which is quoted to be valid (at least) for all
kinds of carbonaceous soot.
In the following, we test the hypothesis of a 
(universal) mass-specific extinction coefficient.
More precisely, we first check if
constant mass-specific extinction does
occur at all in the conducted experiments.
If so, we determine numerical values
of the corresponding extinction coefficients
and compare them with Eq.\,\eqref{eq:av_lita}.
Before starting, it should be noted that
the quantities entering Eq.\,\eqref{eq:alpha_M}
are approximately fixed in our evaluation:
On the one hand, values of $\alpha_{\rm V}$
are obtained from independent MIREX-measurements
and correctly described by the developed model
in terms of $m$ and $\rhop$.
Aerosol densities $\rhoas$, on the
other hand, do not depend on $m$
and are almost independent of the particle
density $\rhop$ which is shown
in Sec.\,\ref{sec:soot_measurements}.
Accordingly, obtained values of $\alpha_{\rm M}$
are approximately a direct output of the
applied experimental and theoretical setup.\\

\subsubsection*{Paraffin aerosol}

\begin{figure}[t!]
    \centering
    \begin{subfigure}{\textwidth}
    \includegraphics[width=.95\linewidth]{figures/%
        paraffin_03_alpha_M_exp19_m=1.42+0.00i_rho=0.85_stokes.png}
    \end{subfigure}
    \caption{
        \textbf{Mass-specific extinction obtained from a paraffin
            ramp experiment using Eq.\,\eqref{eq:alpha_M}.}
        A time average of $\overline{\alpha}_{\rm M}^{}\!=%
        \!\qty{4.32}{\m\squared\per\g}$
        is obtained within the supply-off phase.
        For the model parameters, we use the same values as
        in Fig.\,\ref{fig:paraffin1}, namely
        $m\!=\!1.42$ and $\rhop\!=\!\qty{0.85}{\g\per\cm\cubed}$.
        \label{fig:exp_alphaM_paraffin}}
\end{figure}

To start, we consider a paraffin ramp experiment and 
show results obtained from Eq.\,\eqref{eq:alpha_M}
in Fig.\,\ref{fig:exp_alphaM_paraffin}.
First of all, it is obvious that the value of
$\alpha_{\rm M}(t)$ increases in the supply-on phase
before it stabilizes for times $t\!>\qty{300}{\s}$.
Accordingly, a steady state regarding mass-specific extinction
is indeed forming with an average value in the supply-off phase of
$\overline{\alpha}_{\rm M}^{}\!=\!\qty{4.32}{\m\squared\per\g}$.

Second, in order to check the impact of the particle density
$\rhop$, we consider the average of in total eight experiments
as before and summarize best fit values in Tab.\,%
\ref{tab:parameters_paraffin}.
\begin{table}[ht!]
    \centering
    \caption{
        \textbf{Average $\alpha_{\rm M}$-values obtained
        from paraffin experiments.}
        Shown are results obtained from Eqs.\,\eqref{eq:alpha_M}, 
        \eqref{eq:alpha_Vb}, and \eqref{eq:rho_soot}.
        Given uncertainties of $\overline{\alpha}_{\rm M}$
        correspond to the $3\sigma$-range and include
        variations from one experiment to another
        as well as uncertainties of $\rhop$.        
        \label{tab:parameters_paraffin}}
    \vspace*{-5pt}
    \begin{tabular}{c c c c}
        $\mathfrak{Re}(m)$ &
        $\rhop\,[\unit{\g\per\cm\cubed}]$ &
        \qquad%
        $\overline{\alpha}_{\rm M}^{}\,[\unit{\m\squared\per\g}]$
        \\
        \hline
        $1.30$ & $0.56\pm0.07$ &
        \qquad 
        $4.54\pm 0.92$
        \\
        $1.35$ & $0.68\pm0.07$ &
        \qquad 
        $4.42\pm 0.80$
        \\
        $1.40$ & $0.77\pm0.08$ &
        \qquad 
        $4.38\pm 0.93$
        \\
        $1.45$ & $0.88\pm0.12$ &
        \qquad 
        $4.32\pm 1.08$
        \\
        $1.50$ & $0.99\pm0.12$ &
        \qquad 
        $4.26\pm 0.90$
    \end{tabular}
\end{table}

As expected, the $\rhop$-dependence of
$\overline{\alpha}_{\rm M}$ is only weak.
For instance, a $\qty{77}{\%}$ increase of $\rhop$ in
Tab.\,\ref{tab:parameters_paraffin} corresponds
to an $\overline{\alpha}_{\rm M}^{}$-reduction of only
$\qty{6}{\%}$.
Combining the results, we therefore estimate mass-specific
extinction for a paraffin aerosol at a wave length of
$\lin\!=\!\qty{0.88}{\text{\textmu}\m}$ as
\begin{align}
     \alpha_{\rm M}^{}
     =(4.4\pm1.1)\,\unit{\m\squared\per\g}
     \qquad(99.7\,\%\,\text{c.l.})\,.
     \label{eq:alphaM_paraffin}
\end{align}

\begin{figure}[th!]
    \centering
     \includegraphics[width=.45\textwidth]{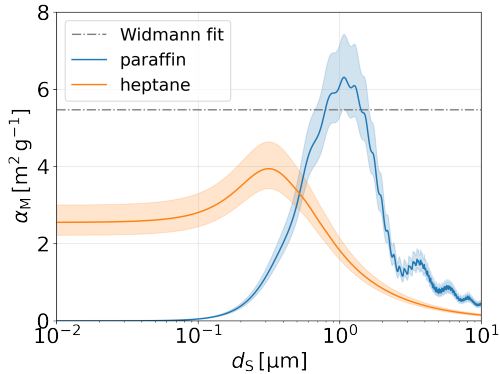}
    \caption{
        \textbf{Mass-specific extinction depending on the
        diameter of the equivalent particle.}
        The curves are obtained from
        Eq.\,\eqref{eq:alpha_Mi} using
        $m\!=\!1.40$,
        $\rhop\!=\!\qty{0.77}{\g\per\cm\cubed}$
        for paraffin and
        $m\!=\!1.57\!+\!0.56\,i$,
        $\rhop\!=\!\qty{2.18}{\g\per\cm\cubed}$
        for soot from a flaming $n$-heptane fire,
        respectively.
        These values are compatible with obtained
        results summarized in Figs.\,%
        \ref{fig:paraffin_parameters} and
        \ref{fig:tf5_parameters_04}.
        The bands indicate a variation of
        $\rhop$ by $\pm\,\qty{15}{\%}$, respectively.
        The wave length of the light is
        $\lin\!=\!\qty{0.88}{\text{\textmu}\m}$.
        \label{fig:alphaMi}}
\end{figure}
Comparing the result with the literature, it is
noticeable that Eq.\,\eqref{eq:alphaM_paraffin}
agrees with Eq.\,\eqref{eq:av_lita} within the
specified uncertainties, but is in general slightly smaller.
This fact, however, is compatible with fundamental
predictions of classical electrodynamics,
which can be seen for instance in Fig.\,\ref{fig:alphaMi}.
Only within a small range of~$\dstokes$ is it
\textit{at all} possible to obtain values equal
to or larger than
$\alpha_{\rm M}\!=\!\qty{5.5}{\m\squared\per\g}$,
e.\,g.\ for $\dstokes\!\approx\!\qty{1}{\text{\textmu}\m}$
(blue curve).
Since the considered aerosol is made of particles
that differ in their individual aerodynamic and
gravimetric  properties (and therefore in
corresponding equivalent diameters~$\dstokes$),
average values of $\alpha_{\rm M}$ can indeed be expected
to be smaller than the \textit{maximum possible}
value of the distribution.
This is true despite the fact that the
$\dstokes$-distribution of a paraffin aerosol is
comparatively narrow, see for instance
the left diagram in 
Fig.\,\ref{fig:exp_alphaV_cum}.

\subsubsection*{Soot from flaming $n$-heptane fire}

To study mass-specific extinction for
soot from a flaming $n$-heptane fire,
we apply the same setup as for the paraffin aerosol.
Corresponding results are summarized in
Tab.\,\ref{tab:parameters_tf5_2}
and Fig.\,\ref{fig:exp_alphaM_41}.
\begin{figure}[t!]
    \centering
    \begin{subfigure}{\textwidth}
        \includegraphics[width=.93\linewidth]{figures/%
            tf5_03_alpha_M_exp4_1_m=1.57+0.56i_rho=2.21_stokes.png}
    \end{subfigure}
    \caption{
        \textbf{Mass-specific extinction from an $n$-heptane
        fire experiment using Eq.\,\eqref{eq:alpha_M}.}
        A time average of
        $\overline{\alpha}_{\rm M}^{}\!=\!\qty{2.15}{\m\squared\per\g}$
        is obtained within the flame-off phase.
        For the model parameters, we use the same values as
        in Fig.\,\ref{fig:exp_alphaV_41}, namely
        $m\!=\!1.57\!+\!0.56\,i$ and
        $\rhop\!=\!\qty{2.21}{\g\per\cm\cubed}$.
        \label{fig:exp_alphaM_41}}
\end{figure}
\begin{table}[ht!]
    \centering
    \caption{
        \textbf{Average $\alpha_{\rm M}$-values obtained
        from $n$-heptane fire experiments.}
        Shown are results obtained from
        Eqs.\,\eqref{eq:alpha_M}, \eqref{eq:alpha_Vb},
        and \eqref{eq:rho_soot}.
        Given uncertainties of $\overline{\alpha}_{\rm M}$
        correspond to the $3\sigma$-range and include
        variations from one experiment to another as well as
        uncertainties of $\rhop$.
        \label{tab:parameters_tf5_2}}
    \vspace*{-15pt}
    \begin{tabular}{ c c c c c }
            $\mathfrak{Re}(m)$ &
            $\mathfrak{Im}(m)$ &
            $\rhop\,[\unit{\g\per\cm\cubed}]$ &
            \qquad
            $\overline{\alpha}_{\rm M}^{}\,[\unit{\m\squared\per\g}]$
            \\
            \hline
            $1.50$   & $0.40$ & $1.80\pm 0.28$ &
            \qquad 
            $2.38\pm 0.96$
            \\
            $1.90$   & $0.40$ & $2.13\pm 0.29$ &
            \qquad 
            $2.38\pm 1.02$
            \\
            \hline
            $1.50$   & $0.50$ & $2.02\pm 0.32$ &
            \qquad 
            $2.37\pm 0.96$
            \\
            $1.90$   & $0.50$ & $2.27\pm 0.31$ &
            \qquad 
            $2.38\pm 0.99$
            \\
            \hline
            $1.57$   & $0.56$ & $2.18\pm 0.35$ &
            \qquad 
            $2.38\pm 1.00$
            \\ 
            \hline
            $1.50$   & $0.60$ & $2.23\pm 0.36$ &
            \qquad 
            $2.38\pm 0.98$
            \\
            $1.90$   & $0.60$ & $2.40\pm 0.34$ &
            \qquad 
            $2.38\pm 1.01$
            \\ 
            \hline
            $1.50$   & $0.70$ & $2.45\pm 0.41$ &
            \qquad 
            $2.38\pm 1.01$
            \\
            $1.90$   & $0.70$ & $2.53\pm 0.37$ &
            \qquad 
            $2.38\pm 1.02$
    \end{tabular}
\end{table}

As for paraffin, a steady state with constant 
average values of $\alpha_{\rm M}$ is forming in Fig.\,%
\ref{fig:exp_alphaM_41}.
Moreover, almost all $\overline{\alpha}_{\rm M}$-values
in Tab.\,\ref{tab:parameters_tf5_2} are identical,
irrespective of the actual values of $m$ and $\rhop$.
We therefore estimate mass-specific extinction
for soot from flaming $n$-heptane fires~as
\begin{align}
    \alpha_{\rm M}
    &=(2.4\pm1.1)\,\unit{\m\squared\per\g}
     \qquad(99.7\,\%\,\text{c.l.})\,,
     \label{eq:alphaM_tf5}
\end{align}
considering again particle sizes between
$\dstokes\!=\!\qty{0.006}{\text{\textmu}\m}$ and
$\dstokes\!\approx\!\qty{4}{\text{\textmu}\m}$
as well as $\lin\!=\!\qty{0.88}{\text{\textmu}\m}$.

In the assessment of the result, the following
points should be noted:
First, the obtained value is significantly
smaller than the value in Eq.\,\eqref{eq:av_lita}
and even outside the provided uncertainties.
Similar discrepancies towards the referenced literature
value, however, are reproduced within the scope of
a study that has been conducted recently~\cite{borger:2022}.
Here, the setup of the experiment was mapped by a numerical
CFD model assuming a mass-specific extinction coefficient
as defined in Eq.\,\eqref{eq:alpha_M}.
Local optical measurements with the MIREX-device revealed
an overestimation of the computed extinction coefficient by
a factor of approximately four.
This result is compatible with Eq.\,\eqref{eq:alphaM_tf5}
if one compares it with the default value of
$ \alpha_{\rm M}\!=\!\qty{8.7}{\m\squared\per\g}$ used in
the simulations \cite{fds, Mulholland_Croarkin_2000}.
On a spatial and temporal scale, the deviation showed an 
almost linear correlation over the entire duration of the
fire. 
Finally, Eq.\,\eqref{eq:alphaM_tf5} is also compatible with
\cite{Wu1997} when using it to compute the corresponding
\textit{dimensionless} extinction coefficient $\K$,
see Tab.\,\ref{tab:parameters2_tf5_2} in Appendix~A2.

From the model point of view, the fact that
the result in Eq.\,\eqref{eq:alphaM_tf5} is smaller
than the one in Eq.\,\eqref{eq:alphaM_paraffin} is
actually compatible with predictions of classical
electrodynamics, see Fig.\,\ref{fig:alphaMi}.
At first glance, it is obvious that the
\textit{maximum possible} value
of $\alpha_{\rm M}$ is considerably smaller
in case of heptane.
Moreover, for larger $\dstokes$,
$\alpha_{\rm M}$ approaches zero more quickly.
Since convolution with the actual size-distribution
of the aerosol particles does not change this fact,
a smaller average of $\alpha_{\rm M}$
is the consequence.
At this point, it should be noted explicitly that the
curve progressions in Fig.\,\ref{fig:alphaMi} are not
a particular feature of the developed model,
but a general prediction of classical electrodynamics.
The observations made are underpinned by the fact
that the average of $\alpha_{\rm M}$ increases
if one only considers \elpi-channels associated with
small and intermediate-size particles.
For instance, considering only the first
ten \elpi-channels (which still provide
\mbox{$\approx\!\qty{80}{\%}$} of total light extinction,
see the r.\,h.\,s.\ of Fig.\,\ref{fig:exp_alphaV_cum})
and using the most commonly used particle density
$\rhop\!=\!\qty{1.86}{\g\per\cm\cubed}$
\cite{Dobbins1994ComparisonOA},
mass-specific extinction increases to
$\alpha_{\rm M}\!=\!(3.8\!\pm\!0.2)\unit{\m\squared\per\g}$.
This value corresponds to a smaller measuring range of the
\elpi-device as it is the average for equivalent particles
up to a size of $\dstokes\!\approx\!\qty{1}{\text{\textmu}\m}$.
In this case, $\alpha_{\rm M}$ is still slightly smaller
than the value in Eq.\,\eqref{eq:av_lita}, but approaches
the order of magnitude of the fit result.


\newpage
\subsection{Light extinction from \elpi-measurements}
\label{sec:application}

\begin{figure}[ht!]
    \centering
    \begin{subfigure}{\textwidth}
        \includegraphics[width=.93\linewidth]{
            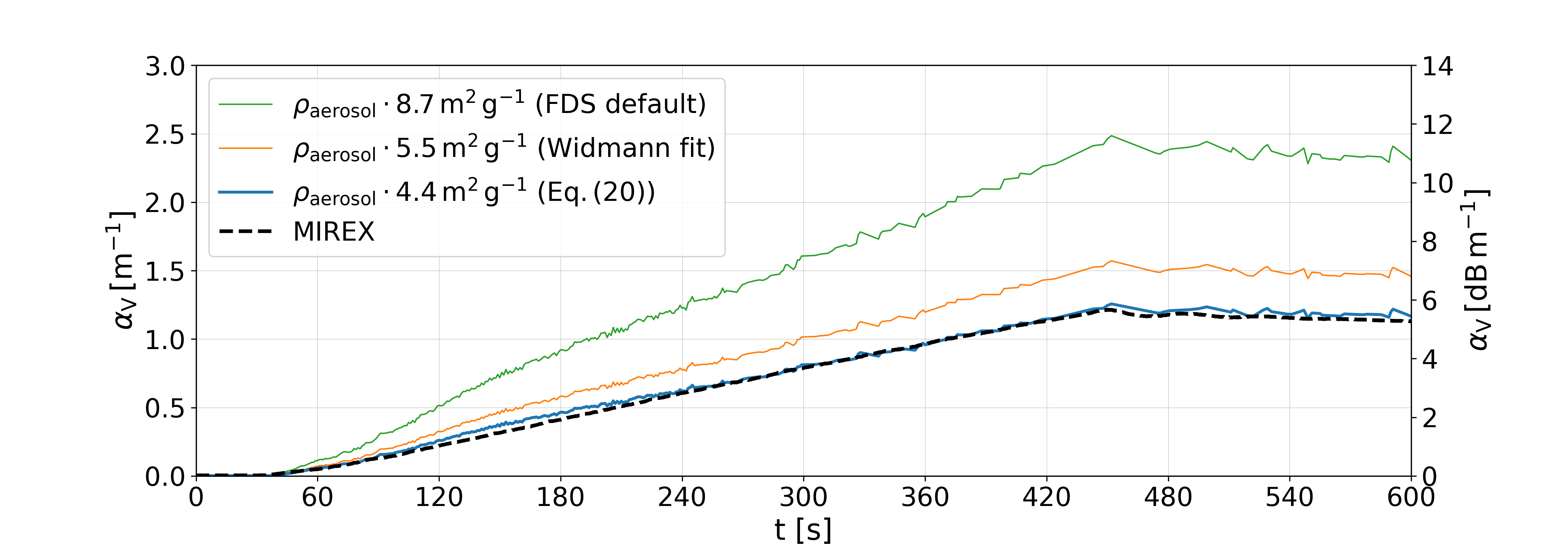}
    \end{subfigure}
    \\[-8pt]
    \begin{subfigure}{\textwidth}
        \includegraphics[width=.93\linewidth]{
            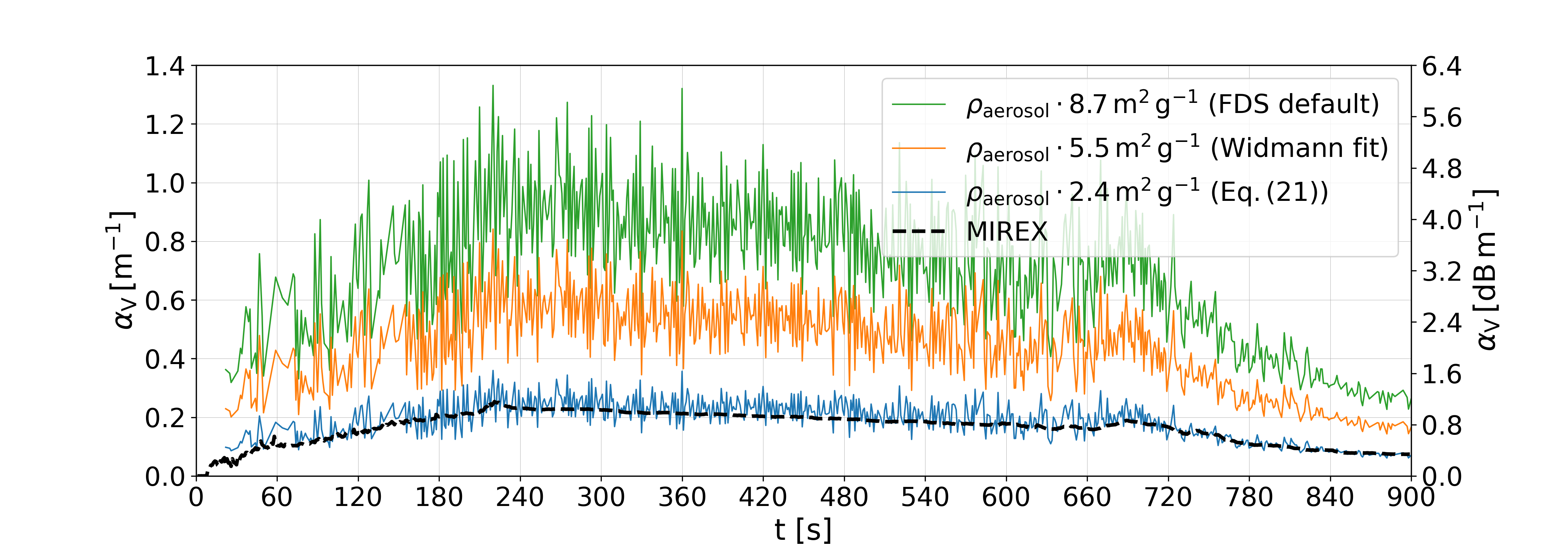}    
    \end{subfigure}
    \caption{
        \textbf{Extinction coefficient $\alpha_{\rm V}$ obtained
        from aerosol density measurements.}
        Shown are results for a paraffin aerosol (upper diagram)
        and soot from a flaming $n$-heptane fire (lower diagram).
        In both diagrams, the wave length of the light
        source is $\qty{0.88}{\text{\textmu}\m}$.
        \label{fig:alpha_rho}}
\end{figure}

Now that values of $\alpha_{\rm M}$ are
obtained for the considered aerosols,
the results can be used to determine visibility
conditions in future applications of the \elpi-system
or measuring devices that work in a similar way.
This can be done by reversing the logic that has
been used to determine $\alpha_{\rm M}$-values in the
first place.
For this, we rewrite Eq.\,\eqref{eq:alpha_M} as
\begin{align}
    \alpha_{\rm V}(t)
    &= \alpha_{\rm M}^{}\cdot\rhoas^{}(t)\,.
    \label{eq:alpha_V_final}  
\end{align}
Accordingly, the only quantity that has to be extracted
from \elpi-data in future applications is the overall
mass density of the aerosol under investigation,
e.\,g.\ via Eq.\,\eqref{eq:rho_soot}.
Multiplication with $\alpha_{\rm M}$
then yields results of $\alpha_{\rm V}$ which can
ultimately be used in the Beer–Lambert–Bouguer law
in Eq.\,\eqref{eq:beer_lambert_bouguer}.
Results obtained using this approach
are shown in Fig.\,\ref{fig:alpha_rho}.
\vspace*{-15pt}

As can be seen, the prediction of
Eq.\,\eqref{eq:alpha_V_final}
agrees much better with MIREX-results when using
$\alpha_{\rm M}$-values obtained in
Eqs.\,\eqref{eq:alphaM_paraffin}
and \eqref{eq:alphaM_tf5} instead of the one in
Eq.\,\eqref{eq:av_lita} or the FDS default value
$\alpha_{\rm M}\!=\!\qty{8.7}{\m\squared\per\g}$
\cite{fds,Mulholland_Croarkin_2000}.
In particular the latter leads to an
overestimation of light extinction of in part more
than a factor of three.

In general, the determination of $\rhoas$ from
\elpi-data is comparatively simple since it is
approximately a direct output of the measuring device,
see e.\,g.\ Eq.\,\eqref{eq:rho_soot} and the discussion
below that equation.
Moreover, due to the fact that $\rhoas$ is almost
independent of the particle density $\rhop$,
it is even possible to assume
$\rhop\!=\!\qty{1}{\g\per\cm\cubed}$ in the
determination of the aerosol density.%
\footnote{
In fact, this corresponds to using aerodynamic
equivalent particles in the determination of 
$\rhoas$.}
Therefore, the actual value of $\rhop$ does
not necessarily have to be known when using 
Eq.\,\eqref{eq:alpha_V_final} to determine
visibility conditions in future applications
of the \elpi-device.\\


\section{Conclusions and outlook}
\label{sec:conclusions}

In this contribution, we develop and apply
a model describing light extinction
in the presence of arbitrary aerosols.
We do this by combining aerosol measurements performed
with the cascade impactor \elpi\ and light-intensity
reduction measurements performed with the extinction
measuring equipment MIREX.
A main motivation in developing the model was that it
should yield reliable predictions of visibility
conditions, but should at the same time be
as simple as possible.
Accordingly, several assumptions were made in order to
focus on most relevant characteristics of the aerosol.
This results in the fact that model predictions
essentially depend on only three independent parameters:
the universal particle density~$\rhop$
as well as real and imaginary parts of the
refractive index~$m$.
Despite the strong restrictions, Eq.\,\eqref{eq:alpha_Vb}
is able to correctly describe time-dependent light
extinction in the presence of an arbitrary aerosol.
This comprises, among other things, different phases
of the aerosol generation (flame-on, flame-off,
exhaust-on phase) as well
as other experimental conditions (bench scale and
room scale, ventilation conditions, etc.).
As it turns out, valid values of the input
parameters $\rhop$, $\mathfrak{Re}(m)$, and
$\mathfrak{Im}(m)$ are simultaneously compatible
with well-established value ranges found in the
literature.

The reason why the model works so well is that
\elpi-measurements simultaneously combine,
through the working principle of the device,
aerodynamic and gravimetric properties of the
collected aerosol particles.
This fact can be capitalized on especially when
Stokes equivalent spheres are used in the evaluation
of the measurement data instead of aerodynamic ones.
Only in the former case, particle properties
are adequately considered and introduced
equivalent spheres have a size compatible with
the size of the actual aerosol particles.
In this way, the combination of \elpi-data and
developed model is able to correctly take into
account large hierarchies of magnitudes which
range for individual cross sections from
$\mathcal{O}(\sigma_{\rm ext})%
\!=\!10^{-22}\,\unit{\m\squared}$
for the smallest particles up to
$\mathcal{O}(\sigma_{\rm ext})%
\!=\!10^{-10}\,\unit{\m\squared}$
for the largest particles.
In doing so, the model gets by without the need of
introducing additional `correction' or `conversion'
factors and correctly predicts extinction coefficients
that are typically in the order
$\mathcal{O}(\alpha_{\rm V})\!=\!\qty{1}{\per\m}$.

As a first application of the developed model,
it is found that main contributions to extinction
stem from only a small number of neighboring
\elpi-channels that correspond to
intermediate-size particles.
More precisely, for a paraffin aerosol, more than
$\qty{90}{\%}$ of extinction originate from only
three \elpi-channels,
whereas for soot from a flaming $n$-heptane fire,
$\approx\qty{80}{\%}$ of extinction
stem from only four channels.
This is one of the main reasons, why the
concept of a \textit{universal} particle density
(which is considered to be the same for each
\elpi-channel) works at all and why
valid numerical values are close to known
literature values.
In a second application, we verify the existence of
constant mass-specific extinction, at least for
the two considered aerosols.
For them, we get the following results which are valid
for the wave length $\lin\!=\!\qty{0.88}{\text{\textmu}\m}$
using the first $13$ channels of the
\elpi-device (particle sizes between
$\dstokes\!=\!\qty{0.006}{\text{\textmu}\m}$ and
$\dstokes\!\approx\!\qty{4}{\text{\textmu}\m}$):
\vspace*{-5pt}
\begin{subequations}
\label{eq:am_final}
\begin{align}
    \text{\textbf{paraffin aerosol}}
    :\qquad
    \alpha_{\rm M}^{}
    &=(4.4\pm 1.1)\,\unit{\m\squared\per\g}
    \qquad(99.7\,\%\,\text{c.l.})\,,\\
    \text{\textbf{soot from flaming $n$-heptane fire}}
    :\qquad
    \alpha_{\rm M}^{}
    &=(2.4\pm 1.1)\,\unit{\m\squared\per\g}
    \qquad(99.7\,\%\,\text{c.l.})\,.
\end{align}
\end{subequations}
The results are not universal in the sense that they
have the same numerical value irrespective of the
present aerosol.
It is important to note, however, that this is not a drawback.
On the contrary, each of the values rather
considers specific aerosol characteristics
that are relevant regarding light extinction,
in particular size-distributions and value ranges
of the equivalent particles.
Therefore, Eqs.\,\eqref{eq:am_final}
can ultimately be used to reliably predict
visibility conditions in future applications
of the \elpi-device.
For this, it is sufficient to determine aerosol densities
$\rhoas^{(13)}$
via Eq.\,\eqref{eq:rho_soot} and to multiply them
with the respective value of~$\alpha_{\rm M}^{}$.
The determination of $\alpha_{\rm M}$-values
describing further aerosols as well as the investigation
of their wave-length dependence are beyond the scope of
the present contribution.
This work is left for future investigations.

\section*{Acknowledgements}
We would like to thank Alica Kandler from the
Institute for Advanced Simulation, Forschungszentrum Jülich, for active
support in the implementation of the aerosol measurements.
Parts of the present contribution are funded by
Deutsche Forschungsgemeinschaft
(DFG, German Research Foundation)
under the project number 465392452.\\

\appendix
\section{Appendix}
\subsection*{A1 \elpi\ conversion factors}
\label{sec:appendix2}

According to \cite{elpi}, \elpi\ conversion factors
are given by
\begin{subequations}
\label{eq:conversion_factor}
\begin{align}
    X_{\rm N_{}}&=
        \frac{1}{1.8300}\cdot
        \frac{\jcal}{\jjet}\cdot
        \xdil\cdot
        \left(\frac{1}{\dstokes}\right)^{1.225}\,,
        \qquad
        \dstokes<\qty{1.035}{\nm}\,,
        \\
    X_{\rm N_{}}&=
        \frac{1}{1.8114}\cdot
        \frac{\jcal}{\jjet}\cdot
        \xdil\cdot
        \left(\frac{1}{\dstokes}\right)^{1.515}\,,
        \qquad
        \qty{1.035}{\nm}<\dstokes<\qty{4.282}{\nm}\,,
        \\
    X_{\rm N_{}}&=
        \frac{1}{3.3868}\cdot
        \frac{\jcal}{\jjet}\cdot
        \xdil\cdot
        \left(\frac{1}{\dstokes}\right)^{1.085}\,,
        \qquad
        \dstokes>\qty{4.282}{\nm}\,,
    \label{eq:conversion_factord}
\end{align}
\end{subequations}
where $\jcal$, $\jjet$, and $\xdil$ denote
the calibration flow rate, the impactor flow rate, 
and the dilution factor that can be adjusted
for each \elpi-measurement, respectively.\\


\subsection*{A2 Dimensionless extinction coefficient $\K$}
\label{sec:appendix_K}

In the already mentioned review \cite{Widmann_2003}
summarizing mass-specific extinction coefficients
obtained in various small-scale-experiments, the
wave-length dependence of $\alpha_{\rm M}$ is approximately
fitted as $\alpha_{\rm M}\!\sim\!\lin^{-1}$.
Assuming such a functional dependence, it is possible to
define a \textit{dimensionless} extinction
coefficient $\K$ in the following way:
\begin{align}
    \K=\alpha_{\rm M}\cdot\rhop\cdot\lin\,.
\end{align}
Written in terms of $\K$, the Beer-Lambert-Bouguer law
in Eq.\,\eqref{eq:beer_lambert_bouguer} is then given by
\begin{align}
    \frac{I(t)}{I_0}
    =\text{exp}\big[-\alpha_{\rm V}\cdot l\big]
    =\text{exp}\big[-\alpha_{\rm M}\cdot\rhoas\cdot l\big]
    =\text{exp}\left[-\K\cdot\frac{\rhoas}{\rhop}\cdot\frac{l}{\lin}\right].
\end{align}
Using results obtained in Secs.\,\ref{sec:ext_coeff} and~%
\ref{sec:specific_extinction}, values of $\K$ for a
paraffin aerosol and soot from a flaming $n$-heptane fire are
summarized in Tabs.\,\ref{tab:parameters2_paraffin} and~%
\ref{tab:parameters2_tf5_2}, respectively.

\begin{table}[ht!]
    \centering
    \caption{
        \textbf{$\K$-values obtained for a paraffin aerosol.}        
        \label{tab:parameters2_paraffin}}
    \vspace*{-5pt}
    \begin{tabular}{c c c c}
        $\mathfrak{Re}(m)$ &
        $\rhop\,[\unit{\g\per\cm\cubed}]$ &
        $\alpha_{\rm M}^{}\,[\unit{\m\squared\per\g}]$ & \qquad
        $\K\,[1]$
        \\
        \hline
        $1.35$ & $0.68$ & $4.42$ & \qquad $2.6$
        \\
        $1.40$ & $0.77$ & $4.38$ & \qquad $3.0$
        \\
        $1.45$ & $0.88$ & $4.32$ & \qquad $3.3$
        \\
        $1.50$ & $0.99$ & $4.26$ & \qquad $3.7$
    \end{tabular}
\end{table}

\begin{table}[ht!]
    \centering
    \caption{
        \textbf{$\K$-values obtained for soot from a flaming
            $n$-heptane fire.}        
        \label{tab:parameters2_tf5_2}}
    \vspace*{-8pt}
    \begin{tabular}{ c c c c c c }
            $\mathfrak{Re}(m)$ &
            $\mathfrak{Im}(m)$ &
            $\rhop\,[\unit{\g\per\cm\cubed}]$ &
            $\alpha_{\rm M}^{}\,[\unit{\m\squared\per\g}]$ &
            \qquad
            $\K\,[1]$
            \\
            \hline
            $1.50$   & $0.40$ & $1.80$ & $2.38$ & \qquad $3.8$
            \\
            $1.90$   & $0.40$ & $2.13$ & $2.39$ & \qquad $4.5$
             \\
            \hline
            $1.50$   & $0.50$ & $2.02$ & $2.37$ & \qquad $4.6$
            \\
            $1.90$   & $0.50$ & $2.27$ & $2.38$ & \qquad $4.8$
            \\
            \hline
            $1.57$   & $0.56$ & $2.18$ & $2.38$ & \qquad $4.6$
            \\ 
            \hline
            $1.50$   & $0.60$ & $2.23$ & $2.38$ & \qquad $4.7$
            \\
            $1.90$   & $0.60$ & $2.40$ & $2.38$ & \qquad $5.0$
            \\ 
            \hline
            $1.50$   & $0.70$ & $2.45$ & $2.38$ & \qquad $5.1$
            \\
            $1.90$   & $0.70$ & $2.53$ & $2.38$ & \qquad $5.3$
    \end{tabular}
\end{table}


\newpage
\subsection*{A3 Aerodynamic equivalent diameters}
\label{sec:appendix3}

In Sec.\,\ref{sec:soot_measurements} it is shown
that Stokes equivalent diameters are better suited
for describing light interaction of aerosol particles
than aerodynamic ones.
In the following, we explicitly verify and quantify
this fact by introducing a corresponding conversion
factor
\begin{align}
    C(\rhop)=\frac%
    {{\alpha}_{\rm V,\,Stokes}(\rhop)}
    {{\alpha}_{\rm V,\,aerodynamic}}
    \equiv
    \frac%
    {{\alpha}_{\rm V,\,Stokes}(\rhop)}
    {{\alpha}_{\rm V,\,Stokes}%
        (\rhop\!=\!\qty{1}{\g\per\cm\cubed})}
    \,.
    \label{eq:C_def}
\end{align}
It is important to notice that,
since Stokes diameters depend on the particle
density $\rhop$, this is of course also the case
for values of $C$.

\subsubsection*{Paraffin  aerosol}

\begin{figure}[ht!]
    \centering
    \includegraphics[width=.75\linewidth]{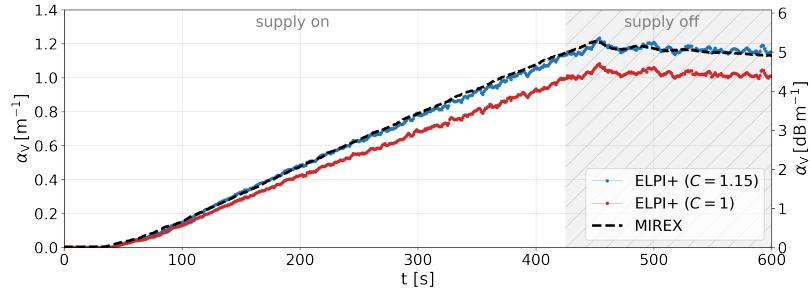}
    \vspace*{-8pt}
    \caption{
        \textbf{Extinction coefficient
        obtained when using aerodynamic equivalent spheres.}
         The value of $m$ is the same as in Fig.\,%
        \ref{fig:paraffin1}.
        The red curve is obtained without correction factor,
        whereas the blue curve is the result of multiplying
        $C\!=\!1.15$.
        \label{fig:paraffin2}}
\end{figure}

Since $\rhop$ is \textit{smaller}
than $\qty{1}{\g\per\cm\cubed}$,
Stokes equivalent diameters are \textit{larger}
than corresponding aerodynamic ones.
For given values of $m$ and $\rhop$,
extinction is therefore smaller when using aerodynamic
equivalent spheres which can be seen 
in Fig.\,\ref{fig:paraffin2}.
\vspace{-15pt}

Agreement with MIREX-results can be obtained
introducing correction factors $C$ as
defined in Eq.\,\eqref{eq:C_def}.
$C$-factors for different values of $m$
(and therefore $\rhop$)
are summarized in Tab.\,\ref{tab:table2}.
\begin{table}[h!]
    \begin{center}
    \caption{
        \textbf{$C$-factors for a paraffin aerosol.}
        \label{tab:table2}
        }
    \vspace*{-8pt}
    \begin{tabular}{c | c c c c c c}
        $\mathfrak{Re}(m)$&
            $1.30$& $1.35$& $1.40$& $1.42$& $1.45$& $1.50$  \\
        $\rhop\,[\unit{\g\per\cm\cubed}]$ &
            $0.58$& $0.70$& $0.80$& $0.85$& $0.92$& $1.03$  \\
        \hline
        $C_{}$&
            $1.70$& $1.41$& $1.21$& $1.15$& $1.07$& $0.98$ 
    \end{tabular}
    \end{center}
\end{table}

\newpage
\subsubsection*{Soot from flaming $n$-heptane fire}

\begin{figure}[ht!]
    \centering
    \includegraphics[width=.85\linewidth]{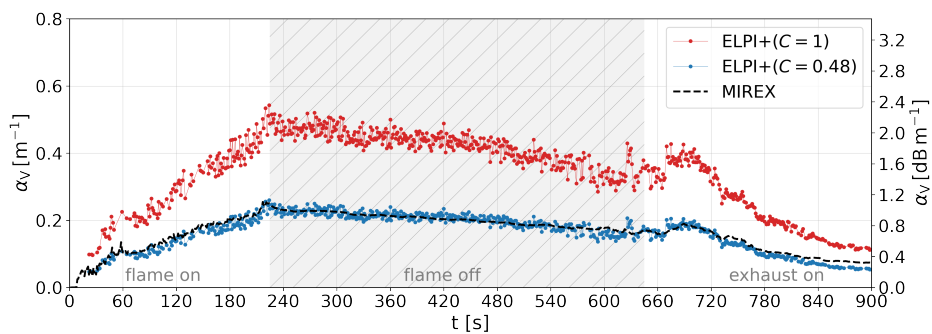}
    \caption{
        \textbf{Extinction coefficient
        obtained when using aerodynamic equivalent spheres.}
        The value of $m$ is the same as in Fig.\,%
        \ref{fig:exp_alphaV_41}.
        The red curve is obtained without correction factor,
        the blue curve is the result of multiplying $C\!=\!0.48$.
        \label{fig:tf5_2}}
\end{figure}

Since $\rhop$ is \textit{larger} than
$\qty{1}{\g\per\cm\cubed}$,
Stokes equivalent diameters are \textit{smaller}
than corresponding aerodynamic ones.
For given values of $\rhop$ and $m$,
extinction is therefore \textit{larger} when using
aerodynamic equivalent spheres which can be seen
in Fig.\,\ref{fig:tf5_2}.
\vspace*{-10pt}

Agreement with the MIREX-measurement can be obtained
when using the overall correction factors summarized
in Tab.\,\ref{tab:parameters_tf5_3}.
\begin{table}[ht!]
    \centering
    \caption{
        \textbf{$C$-factors for soot from a flaming
        $n$-heptane fire.}
        \label{tab:parameters_tf5_3}}
    \vspace*{-8pt}
    \begin{tabular}{ c c c c c }
            $\mathfrak{Re}(m)$ &
            $\mathfrak{Im}(m)$ &
            $\rhop\,[\unit{\g\per\cm\cubed}]$ &
            \qquad $C$\\
            \hline
            $1.50$   & $0.40$ & $1.80$ &\qquad $0.57$
            \\
            $1.90$   & $0.40$ & $2.13$ &\qquad $0.46$
            \\
            \hline
            $1.50$   & $0.50$ & $2.02$ &\qquad $0.52$
            \\
            $1.90$   & $0.50$ & $2.27$ &\qquad $0.44$
            \\
            \hline
            $1.57$   & $0.56$ & $2.18$ &\qquad $0.48$
            \\ 
            \hline
            $1.50$   & $0.60$ & $2.23$ &\qquad $0.48$
            \\
            $1.90$   & $0.60$ & $2.40$ &\qquad $0.42$
            \\ 
            \hline
            $1.50$   & $0.70$ & $2.45$ &\qquad $0.45$
            \\
            $1.90$   & $0.70$ & $2.53$ &\qquad $0.41$
    \end{tabular}
\end{table}

\bibliography{bibliography}{}
\bibliographystyle{JHEP}

\end{document}